%% file: yz.tex
\documentclass[iop,apj]{emulateapj-rtx4}
\usepackage{lscape}
\usepackage{ulem}
\usepackage[backref,breaklinks,colorlinks,citecolor=blue]{hyperref} 
\usepackage[all]{hypcap} 

\begin{document}
\title{X-RAY OUTBURSTS OF LOW-MASS X-RAY BINARY TRANSIENTS OBSERVED IN THE RXTE ERA}
\author{Zhen Yan and Wenfei Yu}
\affil{Key Laboratory for Research in Galaxies and Cosmology, Shanghai Astronomical Observatory, Chinese Academy of Sciences, 80 Nandan Road, Shanghai 200030, China. }
\email{\rm{zyan@shao.ac.cn, wenfei@shao.ac.cn}}

\begin{abstract}
We have performed a statistical study of the properties of 110 bright X-ray outbursts in 36 low-mass X-ray binary transients (LMXBTs) seen with the All-Sky Monitor (2--12 keV) on board the {\it Rossi X-ray Timing Explorer} ({\it RXTE}) in 1996--2011. We have measured a number of outburst properties, including peak X-ray luminosity, rate of change of luminosity on a daily timescale, $e$-folding rise and decay timescales, outburst duration, and total radiated energy. We found that the average  properties such as peak X-ray luminosity, rise and decay timescales, outburst duration, and total radiated energy of black hole LMXBTs, are at least two times larger than those of neutron star LMXBTs, implying that the measurements of these properties may provide preliminary clues as to the nature of the compact object of a newly discovered LMXBT. We also found that the outburst peak X-ray luminosity is correlated with the rate of change of X-ray luminosity in both the rise and the decay phases, which is consistent with our previous studies. Positive correlations between total radiated energy and peak X-ray luminosity, and between total radiated energy and the $e$-folding rise or decay timescale, are also found in the outbursts. These correlations suggest that the mass stored in the disk before an outburst is the primary initial condition that sets up the outburst properties seen later. We also found that the outbursts of two transient stellar-mass ULXs in M31 also roughly follow the correlations, which indicate that the same outburst mechanism works for the brighter outbursts of these two sources in M31 that reached the Eddington luminosity. 
\end{abstract}


\section{INTRODUCTION}
Low-mass X-ray binary transients (LMXBTs) are X-ray transients that contain a primary star of either a black hole (BH) or a neutron star (NS) and a less massive normal star \citep[in general $M \le 1 M_{\sun}$ ][]{liu07}. They are also referred to as X-ray novae (XN) or soft X-ray transients (SXTs) in the literature. LMXBTs are characterized by a long period of quiescence interrupted by episodic outbursts, during which the X-ray flux increases by several orders of magnitude \citep[see][for a review]{tanaka96,chen97}. Outbursts of LMXBTs are usually thought to be triggered by the thermal-viscous instability in a thin accretion disk supposed to exist during the quiescence of LMXBTs \citep[see the reviews by][and references therein]{lasota01}. In the disk instability model (DIM), as matter accumulates in the disk, the surface density increases until a threshold is exceeded at a certain radius, which triggers the instability. The heating front propagates inward and outward, during which the accretion rate increases and the source turns into an outburst phase. However, the original DIM fails to explain some properties of the outbursts in LMXBTs, such as outburst duration, recurrence time, and decay time, so some other processes (such as irradiation) must be considered in the modified DIMs \citep[e.g.][]{king98,dubus01}. 

LMXBTs usually display distinct X-ray spectral states during bright outbursts, whether the compact object is a BH or an NS \citep{yu2003}. The spectral variation during each outburst can be described by the hardness intensity diagram \citep[HID; see ][]{Belloni2010,Munoz-Darias2014}. The tracks along the HID also tightly correlate with the jet activities \citep[see ][]{Fender2004,Fender2009}. Spectral state transitions from the hard state to the soft state usually occur during the rise phase of outbursts \citep[see reviews by][]{remillard06}. It has been found that the luminosity at which the state transition occurs (corresponding to the maximum intensity of the right track on the HID) is approximately proportional to the peak X-ray luminosity of the outburst \citep{yu04,yu07,yd07,yy09,tang11}. The rate of change of the X-ray luminosity during the rise phase of an outburst is also approximately proportional to the outburst peak X-ray luminosity \citep{yy09}. These two empirical correlations strongly suggest that nonstationary accretion dominates the development of the bright hard state and the luminosity of the hard-to-soft state transition. Because of the large range of the accretion rate on a timescale of days to months that an outburst covers, the LMXBTs offer a significant advantage in studying the physics of nonstationary accretion.

\citet{chen97} had performed a comprehensive study of the optical and X-ray properties of LMXBTs before the launch of the {\it Rossi X-ray Timing Explorer} ({\it RXTE}). They collected X-ray light curves of 49 outbursts in 24 LMXBTs observed by Gina, Uhuru, Ariel 5, and Vela 5 B in different energy bands. They found that the LMXB transient X-ray light curves display different morphologies, and they classified the morphologies into five types. Some common parameters were used to describe the properties of the outbursts, such as outburst peak X-ray luminosity, amplitude, rise and decay timescales, outburst duration, and the total radiated energy. Their measurements of outburst parameters are limited by the poor coverage of some light curves.

The observations of outbursts in LMXBTs have been greatly enriched since the launch of the {\it RXTE}. The All-Sky Monitor \citep[ASM;][]{levine96} on board the {\it RXTE} operates in the 2--12 keV range, which provides us with 16 yr long light curves of bright X-ray sources with a quality and coverage better than ever before. It is therefore very necessary to update and summarize the current observations of LMXBTs. In this paper, we present a systematic study of the outburst properties of 110 outbursts in 36 LMXBTs, including measurements of peak X-ray luminosity, rate of change  of luminosity on a daily timescale, $e$-folding rise or decay timescale, outburst duration, and total radiated energy. Notice that these measurements depend on the energy range we investigated -- all of our measurements are based on the {\it RXTE}/ASM (2--12 keV) light curves.

\section{DATA ANALYSIS AND RESULTS}
\subsection{Selection of Bright Outbursts}

\begin{figure*} 
\centering 
\includegraphics[width=\linewidth]{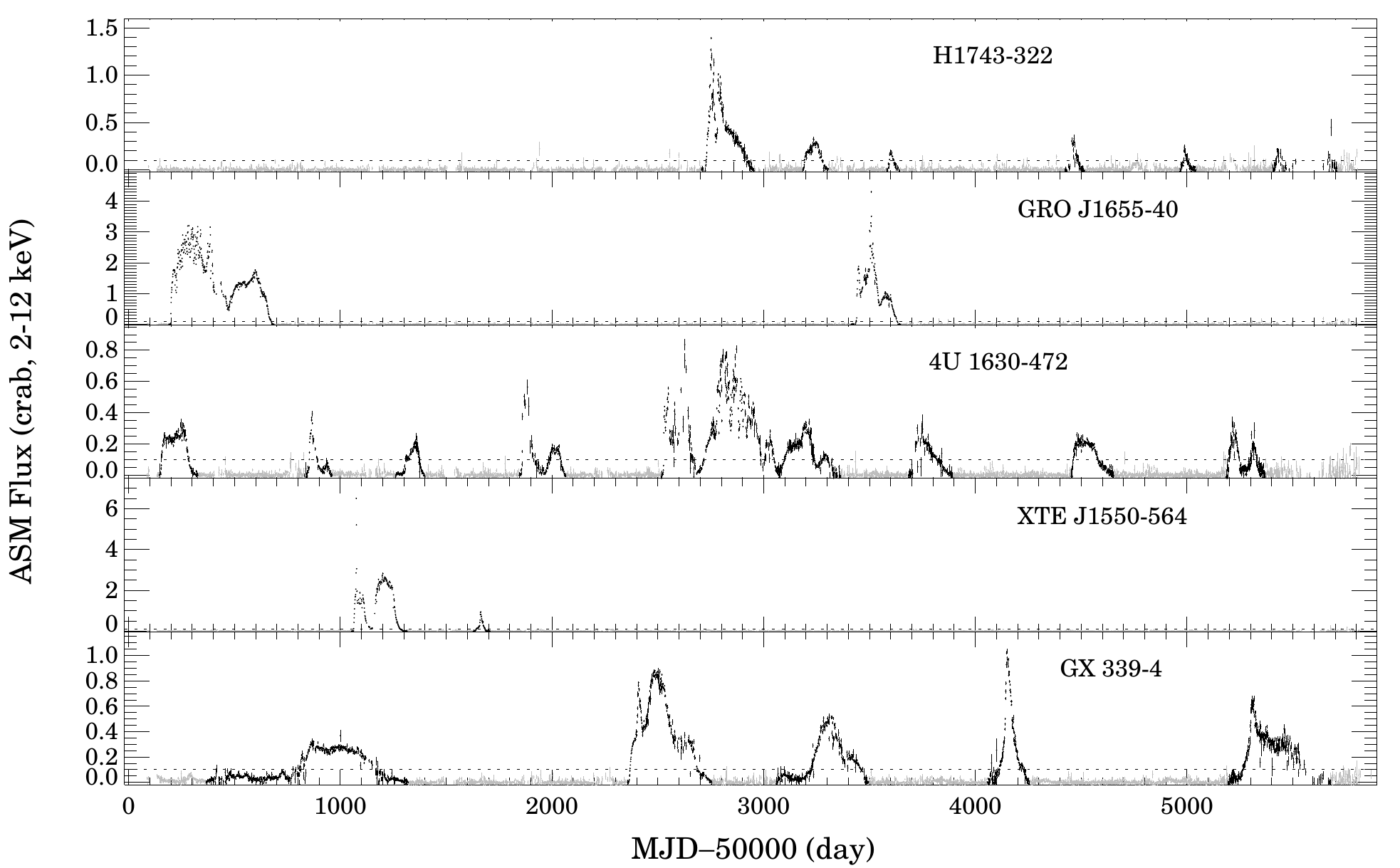}
\includegraphics[width=\linewidth]{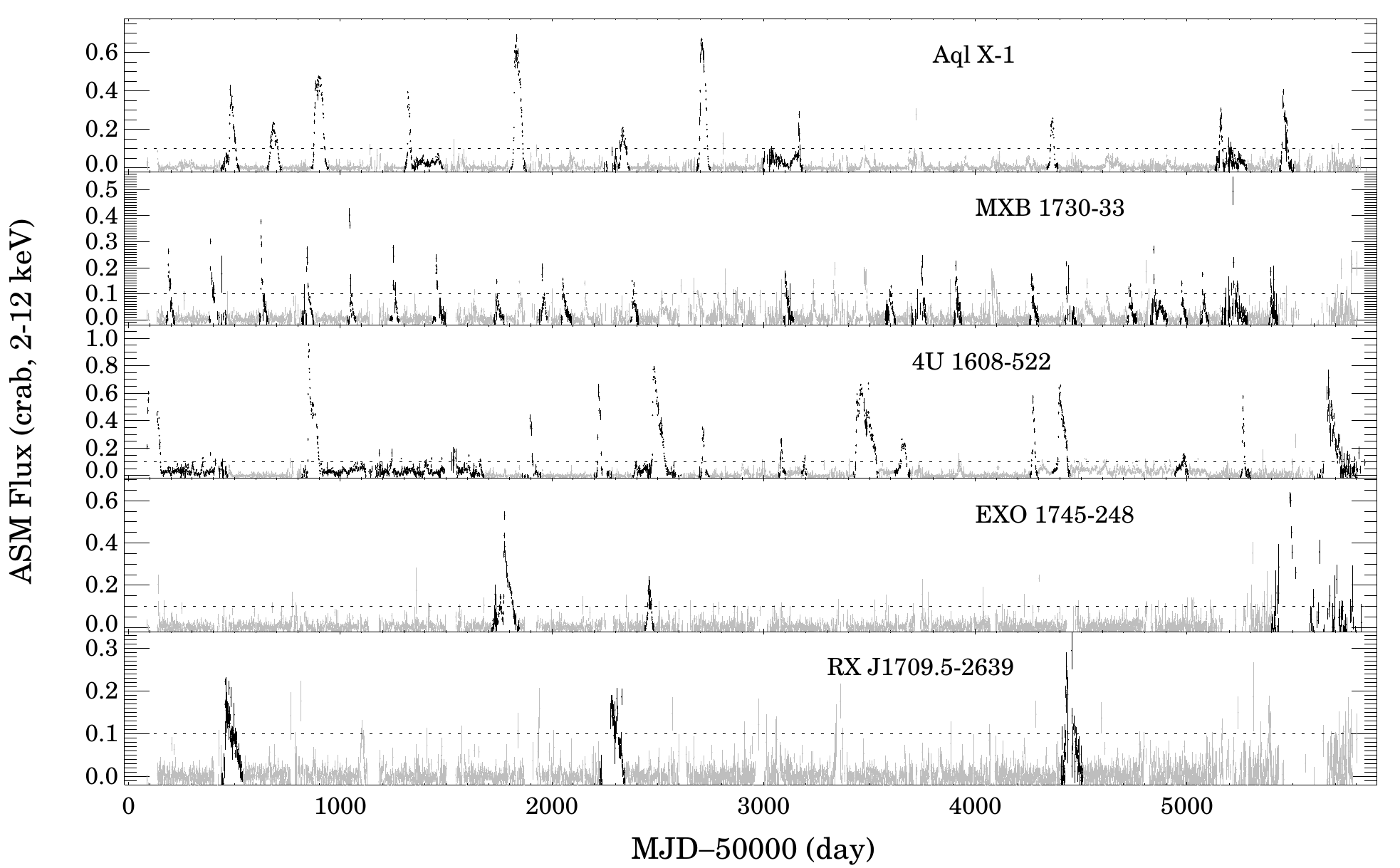}
\caption{Long-term X-ray light curves of LMXBTs with frequent outbursts during 1996--2011 in 2--12 keV. Upper panel: BH LMXBTs. Lower panel: NS LMXBTs. The horizontal dotted line marks the threshold (0.1 Crab) of outburst selection. The light curves of selected outbursts are plotted in black.}
\label{lc}
\end{figure*} 

\begin{figure*}
\centering
\plottwo{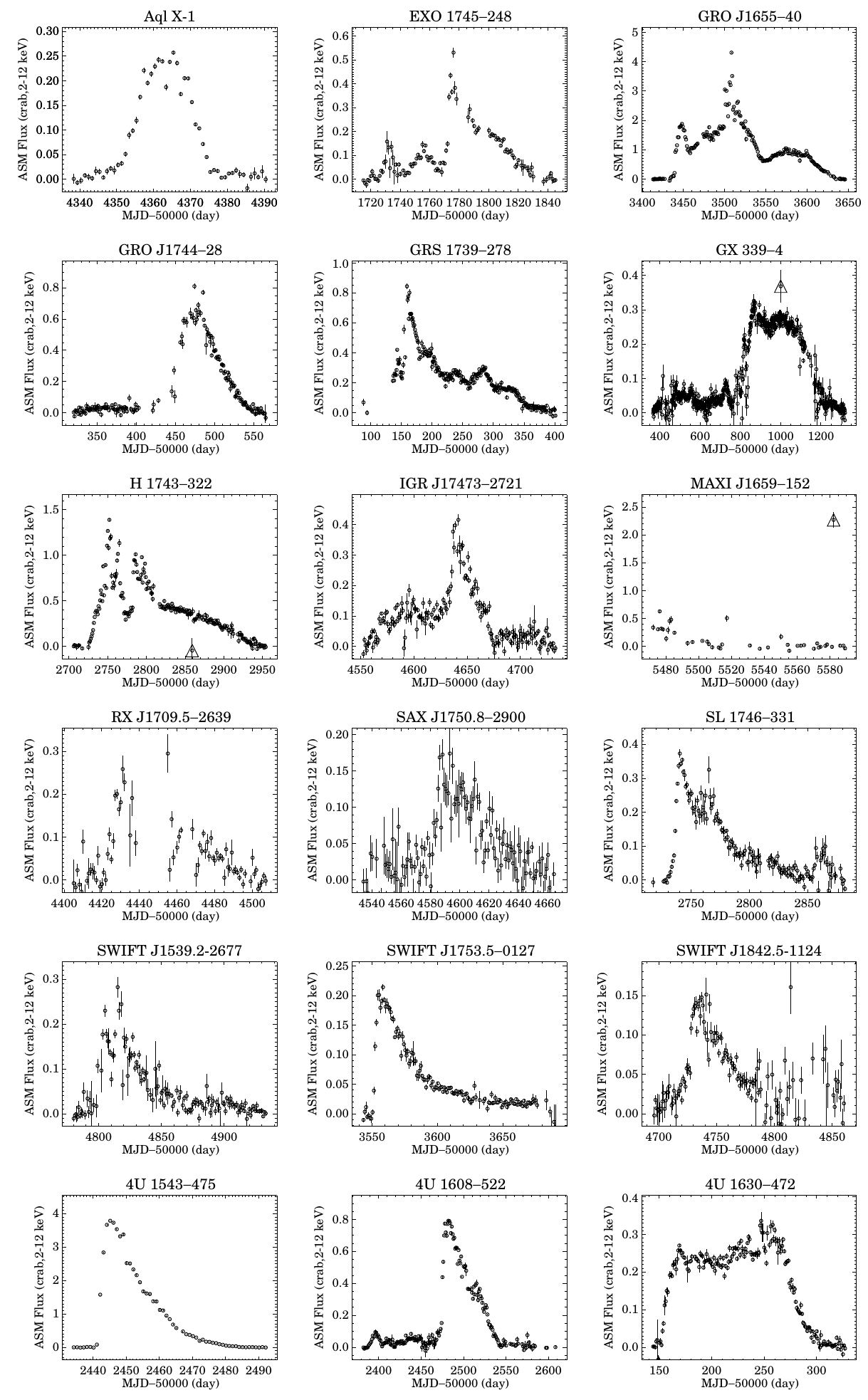}{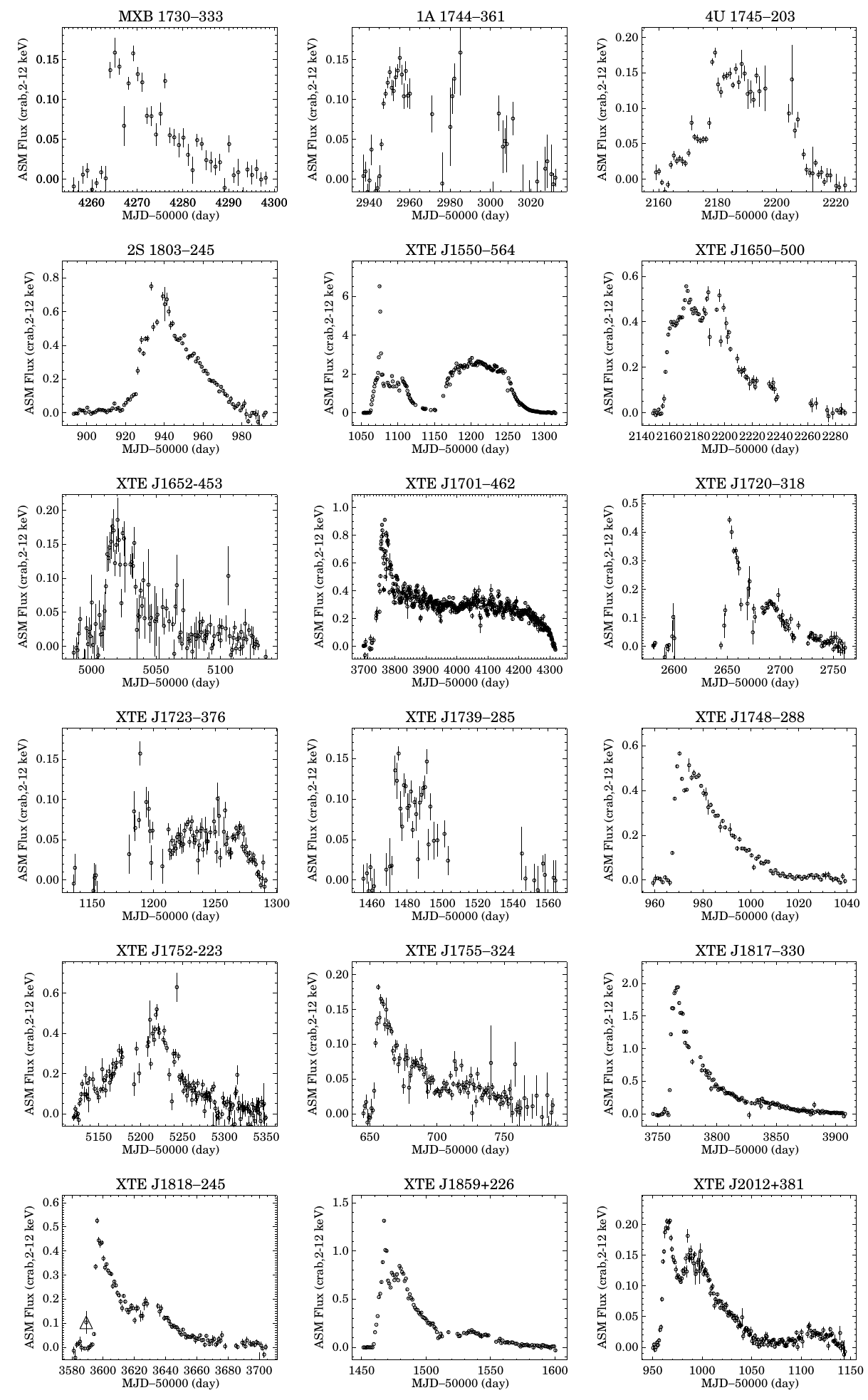}
\caption{\footnotesize Daily light curves of a sample outburst for each LMXBT in our sample in 2--12 keV. The data points plotted as triangle symbols are the examples of data  excluded during the data analysis. See the text of Section of 2.3 and 2.4 for details. }
\label{example1}
\end{figure*}

\capstartfalse
\begin{deluxetable*}{lccccl}
\centering
\tabletypesize{\tiny}
\tablecaption{The System Parameters of LMXBTs in Our Sample}
\tablewidth{0pt}
\tablehead{
\colhead{Source}  & \colhead{$D$($\mathrm{Kpc}$)} &
\colhead{$\mathrm{M_{X} (M_\sun)}$}\tablenotemark{*} & \colhead{$P_\mathrm{orb} (hr)$} &
\colhead{$\mathrm{N_\mathrm{outburst}}$} & \colhead{References}}

\startdata
SWIFT J1539.2$-$6227 &\nodata & \nodata &\nodata & 1 &\citet{krimm11} \\
4U 1543$-$47 & 7.5$\pm$1 & 9.4$\pm$1 & 26.95 &1 &\citet{Orosz2003,orosz98} \\
XTE J1550$-$564 & 5.3$\pm$2.3 & 10.56$\pm$1.0 & 37.25 &2 &\citet{orosz02} \\
4U 1630$-$47 & 10$\pm$5 & \nodata &\nodata &8 &\citet{call00} \\
XTE J1650$-$500 &2.6$\pm$0.7 & 5$\pm$2.3 & 7.63 &1 &\citet{homan06,orosz04} \\
XTE J1652$-$453 &\nodata &\nodata &\nodata & 1 &\citet{mark09}\\
GRO J1655$-$40 & 3.2$\pm$0.2 & 7.02$\pm$0.22 &62.88 & 2 &\citet{orosz97,hjellming95} \\
MAXI J1659$-$152 &7.45$\pm$2.15 &5.8$\pm$2.2 & 2.41 & 1 & \citet{Kuulkers2013,yama12} \\
GX 339$-$4 &5.75$\pm$0.8 & 12.3$\pm$1.4 & 42.14 &5 &\citet{shapo09} \\
XTE J1720$-$318 &6.5$\pm$3.5 &\nodata &\nodata &1 &\citet{chaty06}  \\
GRS 1739$-$278 &7.25$\pm$1.25 &\nodata &\nodata & 1 &\citet{greiner96} \\
H1743$-$322 &9.1$\pm$1.5 &13.3$\pm$3.2 &\nodata &6 &\citet{shapo09} \\
XTE J1748$-$288 & $>$8=10$\pm$2 & \nodata &\nodata &1 &\citet{miller08} \\
SLX 1746$-$331 &\nodata & \nodata &\nodata &3 &\citet{skinner90}\\
XTE J1752$-$223 & 3.5$\pm$0.4 &9.6$\pm$0.9 &\nodata &1 &\citet{sha10} \\
SWIFT J1753.5$-$0127 & 6$\pm$2 & \nodata &3.24 &1 &\citet{cadolle07,zurita08} \\
XTE J1755$-$324 & \nodata & \nodata &\nodata &1 &\citet{rev99} \\
XTE J1817$-$330 & $3\pm2$ & \nodata &\nodata &1 &\citet{sala07} \\
XTE J1818$-$245 & 3.55$\pm$0.75 &\nodata &\nodata &1 & \citet{cb09} \\
SWIFT J1842.5$-$1124 &\nodata &\nodata &\nodata & 1 &\citet{krimm08}\\
XTE J1859$+$226 &4.2$\pm$0.5 &7.7$\pm$1.3 &6.58 &1 &\citet{shapo09,cs11} \\
XTE J2012$+$381 &\nodata &\nodata &\nodata &1 &\citet{camp02} \\
\hline
4U1608$-$522 & 5.8$\pm$2 & \nodata &12.89 &15&\citet{guver10} \\
XTE J1701$-$462 &8.8$\pm$1.3 & \nodata &\nodata &1 &\citet{lin09} \\
RX J1709.5$-$2639 & 10$\pm$2 & \nodata &\nodata &3 &\citet{jonker04} \\
XTE J1723$-$376 &$<$13=10$\pm$3 & \nodata & \nodata &1 &\citet{galloway08} \\ 
MXB 1730$-$33 & 8.8$\pm$3 & \nodata &\nodata &23 &\citet{kuu03} \\
XTE J1739$-$285 & $<$10.6=8$\pm$2 &\nodata &\nodata &1 &\citet{kaaret07} \\
GRO J1744$-$28 & 8.5$\pm$0.5 & \nodata & 284.02 &2 & \citet{nishiuchi99,finger96} \\
IGR J17473$-$2721 &6.4$\pm$0.96 &\nodata &\nodata &2 &\citet{chen10} \\
EXO 1745$-$248 &8.7$\pm$3 &\nodata &\nodata & 3 & \citet{cohn02,kuu03} \\ 
1A 1744$-$361 &$<$9=6$\pm$3 & \nodata & 1.62 &1 &\citet{bhattacharyya06} \\
4U 1745$-$203 & 8.47$\pm$0.4 &\nodata& \nodata &2 & \citet{ortolani94} \\
SAX J1750.8$-$2900 & 6.79 $\pm$0.14 &\nodata &\nodata &2 &\citet{kaaret02,galloway08} \\
2S 1803$-$245 &$<$7.3=5$\pm$2 & \nodata & 9 &1 & \citet{cornelisse07} \\
Aql X-1 & 5$\pm$1 & \nodata & 18.95 &11 & \citet{rutledge01,welsh00} \\
\enddata
\label{parameters}
\tablenotetext{*}{We assumed the NS mass to be 1.4$\pm$0.1 ${M_\sun}$ }
\end{deluxetable*}
\capstarttrue

Among 187 LMXBs in the catalog of \citet{liu07}, 103 sources are identified as transients, 95 of which are monitored by the {\it RXTE}/ASM. We also found 11 LMXBTs in the ASM data archive which are not included in the LMXB catalog of \citet{liu07}. Then there are a total of 106 LMXBTs with {\it RXTE}/ASM light curves, but four of them (GRS 1741.9$-$2853, AX J1745.6$-$2901, 1A 1742$-$289, CXOGC J174540.0$-$290031 ) are in the Galactic center region, which cannot be resolved by the {\it RXTE}/ASM. So we excluded these four sources in our analysis. There are another three sources (GRS 1915$+$105, KS 1731$-$260, and Cir X-1) that have outbursts of very long duration during the $RXTE$ era. The rise or decay phases of the outbursts of these three sources are not available during the {\it RXTE} era, so we do not take these three sources as classic transient sources on the observational timescale of the {\it RXTE}/ASM. IGR J17091$-$3624 is also excluded because of its very poor data coverage. 

Thus a total of 98 LMXBTs with available {\it RXTE}/ASM data are included in our analysis. Their long-term X-ray behavior during the {\it RXTE} era roughly satisfied the criteria for XN in \citet{tanaka96}.  The one-dwell light curve of each source retrieved from the public ASM products database, \footnote{\url{http://heasarc.nasa.gov/docs/xte/asm\_products.html}} which covers the period from early 1996 to late 2011, was uniformly rebinned with one day duration bins  in order to improve the sensitivity ($\sim$ 10 mCrab for the daily average data), and the effects of the Type I and Type II X-ray bursts in NS LMXBTs are insignificant in the one-day binned data. The X-ray flux of each source is reported in Crab units ($1~\mathrm{Crab} = 75~\mathrm{~ct~s}^{-1}$, estimated from the light curve of the Crab Nebula observed by the ASM). We only selected as our samples outbursts with X-ray peak fluxes larger than 0.1 Crab over a significance of $3\sigma$, so our samples only include bright outbursts. We describe how to select the outburst peak flux in Section 2.3 in detail. 
 
The object 4U 1608$-$522 sometimes stayed in a low flux level for as long as about 80 days after an outburst, then showed a secondary outburst, such as the 2004 and 2005 outbursts (during MJD 53072--53202 and MJD 53424--53691, respectively). We took the primary and the secondary outbursts as two independent samples in our analysis. The 2007--2009 outburst of 4U 1608$-$522 \citep[see, e.g.,][or \autoref{lc} during about MJD 54258--55006]{linares09} is constituted of two primary outbursts and and a series of small outbursts. These series of outbursts contributed three outburst samples in our analysis.

Finally, we ended up with a total of 110 outbursts in 36 LMXBTs, which includes 22 BH systems and 14 NS systems, respectively. More than half of the LMXBTs in our sample only have one bright outburst during the $RXTE$ era (see \autoref{parameters}). MXB 1730$-$33 has the most amount of bright outbursts, up to 23 in the 16 yr. System parameters collected from the literature for these LMXBTs are shown in \autoref{parameters}. \autoref{lc} shows long-term X-ray light curves of five NS and BH LMXBTs with the frequent outbursts during 1996--2011. We measured the peak X-ray luminosity, rate of change  of luminosity on a daily timescale, $e$-folding rise or decay timescale, outburst duration and total radiated energy for each of the outbursts. All of the results are collected in \autoref{results}. We plot one outburst from each LMXBT in \autoref{example1} as examples.

\subsection{Bright Outburst Rate}
\begin{figure*} 
\centering 
\includegraphics[width=\linewidth]{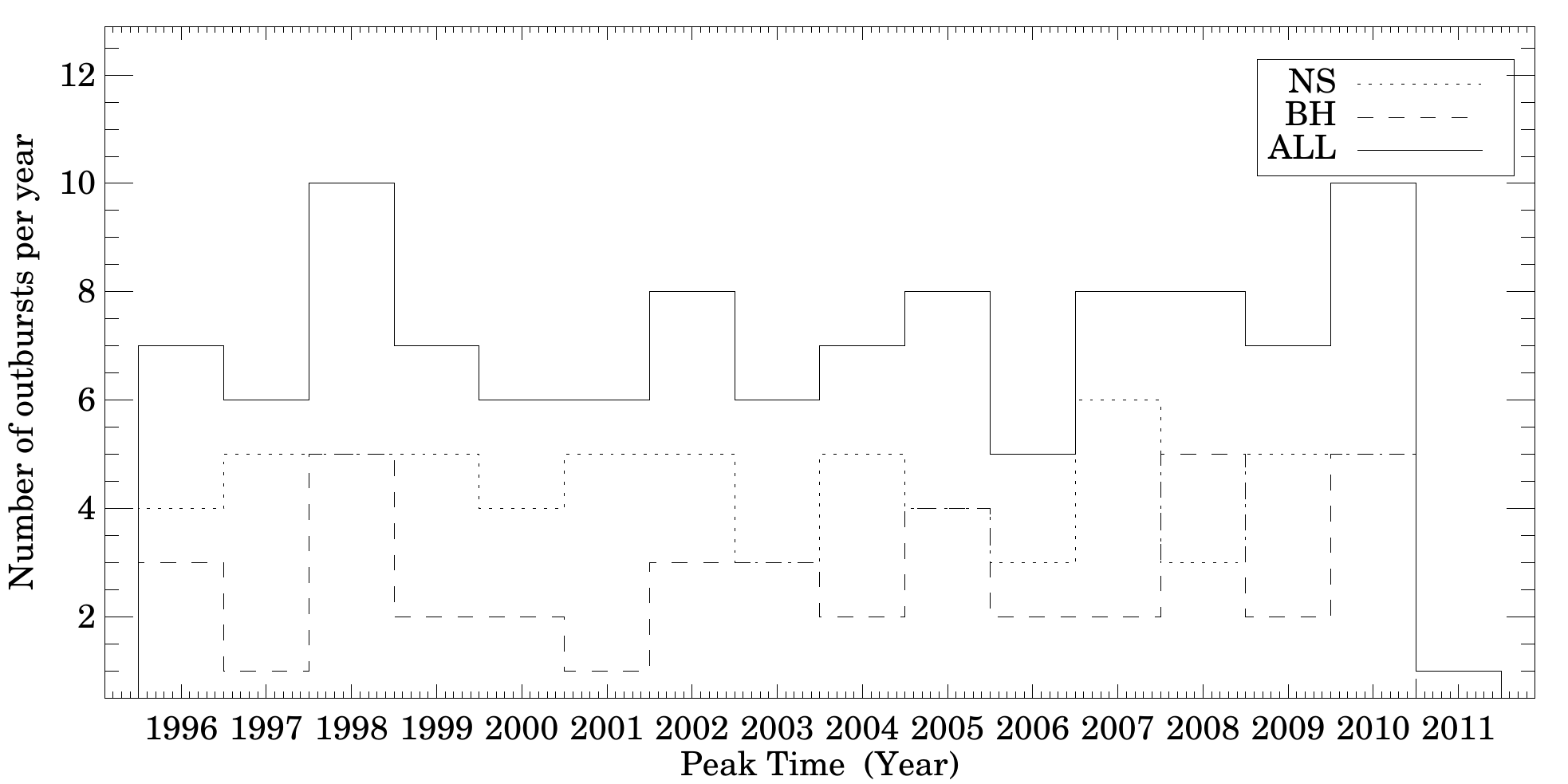}
\caption{The number of Galactic LMXBT bright outbursts ($F_\mathrm{peak} > 0.1$ Crab in 2--12 keV) per year from early 1996 to late 2011. The annual average (excluding the year 2011 ) of bright outbursts is 7.3, 4.5, and 2.8 for all of the LMXBTs, NSs, and BHs, respectively.}
\label{ptime}
\end{figure*}

We identified 110 bright LMXBT outbursts in the observations of {\it RXTE}/ASM (including 68 NS LMXBT outbursts and 42 BH LMXBT outbursts). The numbers of outbursts in every year is shown in \autoref{ptime}. Because of the detector problems of {\it RXTE}/ASM since late 2010, the sky coverage of {\it RXTE}/ASM was highly reduced (A. Levine 2012, private communication), so there is only one outburst (the 2011 outburst of 4U 1608$-$52) can be identified in the 2011 data. For the period of 15 yr from early 1996 to late 2010, an annual average of 7.3 bright outburst events of LMXBTs were detected by {\it RXTE}/ASM, including 4.5 NS LMXBT outbursts and 2.8 BH LMXBT outbursts. This average number of outbursts per year is roughly three times larger than the estimation in \citet{chen97}, which may be due to the improvement of the sky coverage of the {\it RXTE}/ASM. Our outburst sample consists of bright outbursts with a peak flux larger than 0.1 Crab. Therefore, the outburst rate of the LMXBTs we report here is independent of {\it RXTE}/ASM sensitivity. 

 \subsection{Peak X-ray Luminosity}
 \capstartfalse
 \begin{deluxetable*}{lccc}
\centering
\tablecolumns{5}
\tabletypesize{\tiny}
\tablecaption{Statistical Results of the Average Peak X-ray Luminosity}
\tablewidth{0pt}
\tablehead{
\colhead { } & \colhead{All LMXBTs}  & \colhead{NS LMXBTs} &\colhead{BH LMXBTs}  
}

\startdata
$<\log(F_\mathrm{peak})>$ & $-0.440\pm 0.348$ & $-0.531\pm 0.266$ & $-0.292\pm 0.413$ \\
$<F_\mathrm{peak}> (Crab,2-12~keV)$ & $ 0.363^{+ 0.447}_{- 0.200}$ & $ 0.294^{+ 0.249}_{- 0.135}$ &$0.510^{+ 0.812}_{- 0.313}$ \\
$<\log(L_\mathrm{peak})>$ & $37.668\pm 0.331$ & $37.579\pm 0.267$ & $37.846\pm 0.375$ \\
$<L_\mathrm{peak}> (10^{38}~\mathrm{ergs}~\mathrm{s}^{-1},2-12~keV)$ & $ 0.466^{+ 0.531}_{- 0.248}$ & $0.380^{+ 0.322}_{- 0.174}$ &$0.701^{+ 0.964}_{- 0.406}$ \\
$<\log(L_\mathrm{peak}/L_\mathrm{Edd})>$ & $-0.817\pm 0.398$ & $-0.681\pm 0.267$ & $-1.281\pm 0.423$ \\
$<L_\mathrm{peak}/L_\mathrm{Edd}> (2-12~keV)$ & $0.152^{+ 0.228}_{- 0.091}$ & $ 0.209^{+ 0.177}_{- 0.096}$ & $0.052^{+ 0.086}_{- 0.033}$ \\
\enddata
\label{lpeak}
\end{deluxetable*}
\capstarttrue
 
\begin{figure*} 
\centering 
\includegraphics[width=\linewidth]{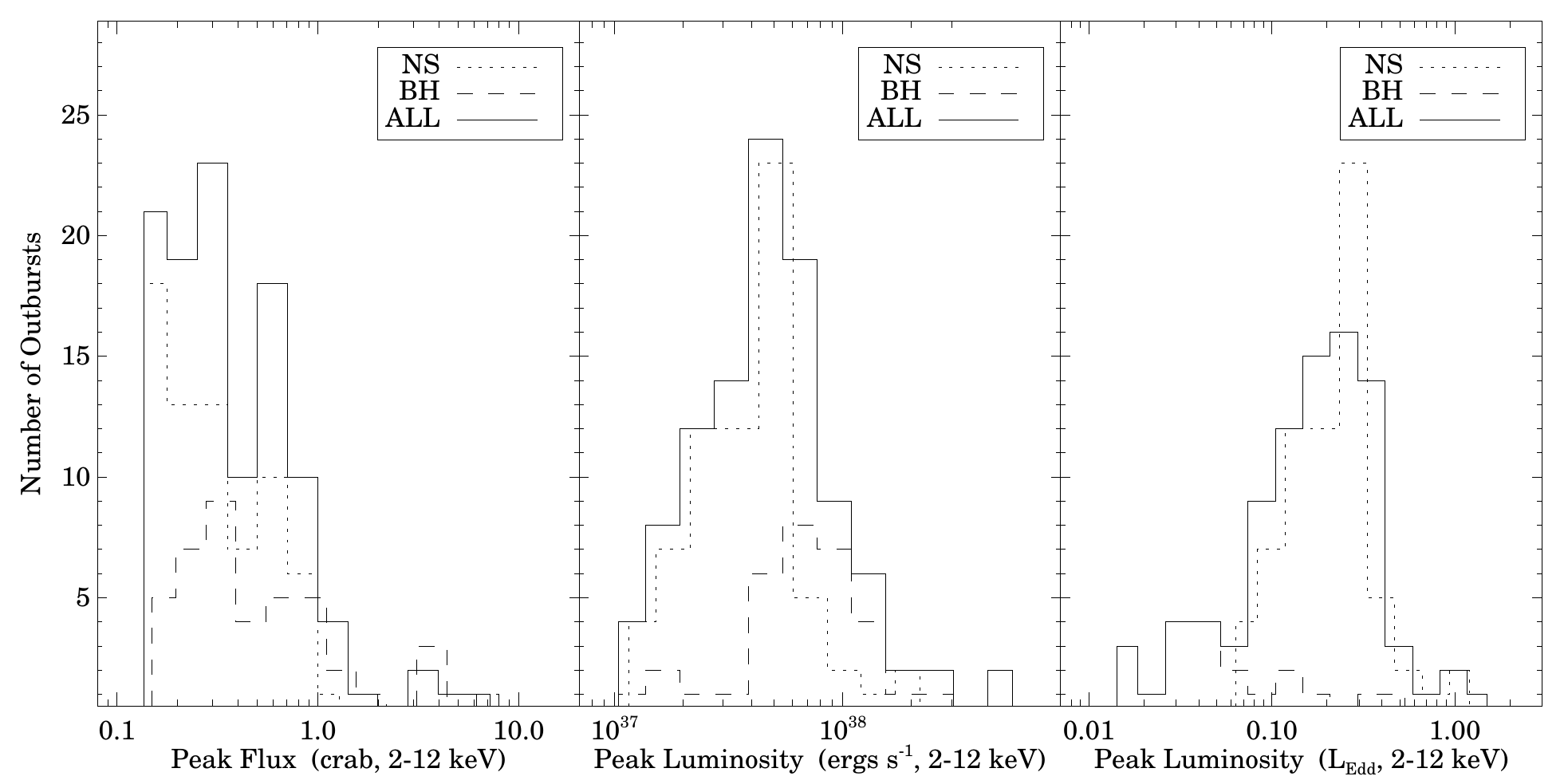}
\caption{The distributions of peak luminosities and fluxes of LMXBT outbursts in 2--12 keV. The left panel shows the peak fluxes in Crab units, of which the average is 0.363 Crab in a logarithmic scale. The middle panel shows the peak luminosities in units of $~\mathrm{erg}~\mathrm{s}^{-1}$, of which the average is 0.466 $\times 10^{38}~\mathrm{erg}~\mathrm{s}^{-1}$ in a logarithmic scale. The right panel shows the peak luminosities in units of $L_\mathrm{Edd}$, of which the average is 0.152 $L_\mathrm{Edd}$ in a logarithmic scale.}
\label{peak}
\end{figure*} 

We chose the maximal flux value in the daily light curve of an outburst as the peak flux. We excluded two kinds of data points when we identified the peak flux of an outburst, which are unlikely to be real detections and may be from instrumental or systematic errors. The first kind is that the flux is larger than the fluxes on two adjacent days by more than a factor of two (e.g., maximum fluxes of the outbursts of RX J1709.5$-$2639 and MAXI J1659$-$152 in \autoref{example1}). The second kind is that the error of the flux is larger than the two adjacent days by more than a factor of two (e.g., maximum flux of an outburst of GX 339$-$4 in \autoref{example1}). In addition, the ASM data of two outbursts have large data gaps near the outburst peaks (the 1996 outburst of 4U 1608$-$522 and the 1996 outburst of GRO J1744$-$28). In such cases, the peak flux we measured may be the lower limit of the true value.

In order to convert the observed photon count rate into flux, we assumed that the X-ray spectrum of each LMXBT is the same as the Crab Nebula in the 2--12 keV energy band. For simplicity, we used a conversion of 1 Crab as $2.12 \times10^{-8}$~ergs~cm$^{-2}$~s$^{-1}$ (2--12 keV), which is estimated according to the spectrum of the Crab Nebula in the 2--10 keV energy band \citep{kirsch05}. The column densities of the sources in our sample are in the range of $0.15-8\times10^{22}$ cm$^{-2}$, and most of them (29 out of 36) are below $2\times10^{22}$ cm$^{-2}$. We have used WebPIMMS to estimate the flux of the spectra of typical soft and hard states of LMXBTs at a column density of $2\times10^{22}$ cm$^{-2}$ with an input $RXTE$/ASM count rate of the Crab Nebula ($\sim$75 c s$^{-1}$), and we found that the differences from the flux of the Crab Nebula are about 10\% and 25\% in the 2--12 keV energy band for the soft sate and the hard state, respectively. So for most sources, the uncertainty of the flux we estimated according to the Crab Nebula spectra is less than 25\%. However, for the large column density cases (such as 4U 1630$-$472, $\sim$ 8$\times10^{22}$ cm$^{-2}$), we may underestimate the flux by 90\% in the 2--12 keV energy band. We collected the values of distances and masses of our selected LMXBTs from literature in order to estimate the X-ray luminosity (the NS mass is assumed to be 1.4 $\pm 0.1 M_{\sun}$). Then the peak X-ray luminosity $L_\mathrm{peak}$ of the sources (14 NS LMXBTs and 9 BH LMXBTs) with known distances and masses are scaled in units of $L_\mathrm{Edd}$, where $L_\mathrm{Edd}$ is taken as $1.3\times 10^{38}(M/M_{\sun})~\mathrm{ergs}~\mathrm{s}^{-1}$. These estimations could bring large uncertainties due to the differences in spectral shapes, hydrogen column density, inclination angle, and radiation efficiency between those of the Crab Nebula and those of the sources in our sample, and in addition to the distances and the masses \citep{zand07}. Notice that the X-ray flux is measured in 2--12 keV. The bolometric X-ray luminosity could be 3 times larger than the luminosity in 2--12 keV including the uncertainties caused by the diverse energy spectra and column densities \citep{zand07}. 

Among the 110 outbursts we selected, the brightest outburst during the $RXTE$ era is the 1998 outburst of XTE J1550$-$564, which reached up to 6.5 Crab. The faintest outburst we selected is the 2004 outburst of 4U 1608$-$522, the peak X-ray flux of which is about 0.12 Crab. The most luminous outburst during the $RXTE$ era is the 2002 outburst of 4U 1543$-$475 for the sources with known distances, which reached to $5.4 \times 10^{38} ~\mathrm{ergs}~\mathrm{s}^{-1}$ in 2--12 keV, and the bolometric X-ray luminosity could be $1.6 \times 10^{39} \mathrm{ergs}~\mathrm{s}^{-1}$. The minimum peak X-ray luminosity of our selected outbursts is $9.6 \times 10^{36} ~\mathrm{ergs}~\mathrm{s}^{-1}$ from the 2001 outburst of XTE J1650$-$500. The peak X-ray luminosity of the 1996 outburst of GRO J1744$-$28 is about 1.3 $L_\mathrm{Edd}$ in 2--12 keV, which is the highest Eddington luminosity in our sample. But $RXTE$/ASM only covered the decay phase of this outburst, so the peak X-ray luminosity we measured could be the lower limit of the true value. The bolometric X-ray luminosity is about three times larger than that in 2--12 keV \citep[e.g.][]{zand07}, so the actual peak X-ray luminosity of  the 1996 outburst of GRO J1744$-$28 is at least 3.9 $L_\mathrm{Edd}$. The minimum peak X-ray in our sample is about 0.01 $L_\mathrm{Edd}$ in 2--12 keV, which is from the 2010 outburst of XTE J1752$-$223.

\autoref{peak} shows the distributions of the peak luminosities and fluxes of all of the selected outbursts, including the BH LMXBTs, the NS LMXBTs and all of the LMXBTs, respectively. The statistical results of the average peak X-ray luminosity in a logarithmic scale are shown in \autoref{lpeak}. The distribution of the peak X-ray fluxes in Crab units has a cutoff at the low flux end due to the selection criteria of 0.1 Crab, and the distribution above 0.1 Crab is roughly consistent with a power-law form. From the left panel of \autoref{peak}, we can see that the outbursts with peak flux larger than 1.5 Crab all belong to BH LMXBTs. The distribution of peak luminosities in units of $\mathrm{ergs}~\mathrm{s}^{-1}$ and $L_\mathrm{Edd}$ both are roughly consistent with a Gaussian-like distribution in a logarithmic scale. The average peak X-ray luminosity (in units of ergs s$^{-1}$) of BH LMXBTs is nearly two times larger than that of NS LMXBTs; we can see those from both the middle panel of \autoref{peak} and \autoref{lpeak}. The average peak X-ray luminosity (in units of $L_\mathrm{Edd}$) of the NS LMXBTs is four times larger than that of the BH LMXBTs (see the right panel of \autoref{peak} and \autoref{lpeak}). The average BH mass with a mass determination in our sample is about 9 $M_{\sun}$, and the NS mass is assumed as 1.4 $M_{\sun}$. By considering the difference in average peak X-ray luminosity in units of $\mathrm{ergs}~\mathrm{s}^{-1}$ between BH and NS LMXBTs, we obtain a peak X-ray luminosity in units of $L_\mathrm{Edd}$ of NS LMXBTs that is about four times larger than that of BH LMXBTs, which is consistent with our measurement.

\subsection{$e$-Folding Timescale and Rate of Change  of Luminosity during the Rise or Decay Phase}
\capstartfalse
\begin{deluxetable*}{lccc}
\centering
\tablecolumns{5}
\tabletypesize{\footnotesize}
\tablecaption{Statistical Results of the Average $e$-Folding Rise and Decay Timescales}
\tablewidth{0pt}
\tablehead{
\colhead { } & \colhead{All LMXBTs}  & \colhead{NS LMXBTs} &\colhead{BH LMXBTs}  
}

\startdata
$<\log(\tau_\mathrm{rise,10\%-90\%})>$ & $0.726\pm 0.477$ & $0.610\pm 0.432 $ & $0.913\pm 0.493$ \\
$<\tau_\mathrm{rise,10\%-90\%}> (\rm days)$ & $ 5.326^{+10.663}_{- 3.552}$ & $4.072^{+ 6.926}_{- 2.564}$ &$8.185^{+17.288}_{- 5.555}$ \\
$<\log(\tau_\mathrm{rise,10\%-50\%})>$ & $0.565\pm 0.516$ & $0.509\pm 0.523$ & $0.639\pm 0.504$ \\
$<\tau_\mathrm{rise,10\%-50\%}> (\rm days)$ & $3.672^{+ 8.377}_{- 2.553}$ & $3.230^{+ 7.533}_{- 2.261}$ & $4.356^{+ 9.552}_{- 2.992}$ \\
$<\log(\tau_\mathrm{rise,50\%-90\%})>$ & $0.847\pm 0.509$ & $0.726\pm 0.415 $ & $1.007\pm 0.578$ \\
$<\tau_\mathrm{rise,50\%-90\%}> (\rm days)$ & $7.038^{+15.677}_{- 4.857}$ & $5.323^{+ 8.509}_{- 3.275}$ &$10.166^{+28.349}_{- 7.483}$ \\
\hline
$<\log(\tau_\mathrm{decay,10\%-90\%})>$ & $1.176\pm 0.392$ & $1.032\pm 0.348$ & $1.410\pm 0.347$ \\
$<\tau_\mathrm{decay,10\%-90\%}> (\rm days)$ & $ 15.013^{+22.022}_{- 8.927}$ & $10.774^{+13.260}_{- 5.944}$ &$25.689^{+31.393}_{-14.128}$ \\
$<\log(\tau_\mathrm{decay,10\%-50\%})>$ & $1.086\pm 0.414$ & $0.925\pm 0.380$ & $ 1.339\pm 0.332$ \\
$<\tau_\mathrm{decay,10\%-50\%}> (\rm days)$ & $12.194^{+19.415}_{- 7.490}$ & $ 8.418^{+11.773}_{- 4.908}$ & $21.829^{+25.098}_{-11.675}$ \\
$<\log(\tau_\mathrm{decay,50\%-90\%})>$ & $1.186\pm 0.490$ & $1.050\pm 0.462 $ & $1.398\pm 0.461$ \\
$<\tau_\mathrm{decay,50\%-90\%}> (\rm days)$ & $15.330^{+32.075}_{-10.373}$ & $11.221^{+21.291}_{- 7.348}$ &$25.032^{+47.397}_{-16.381}$ \\
\enddata
\label{tau}
\end{deluxetable*}
\capstarttrue
 
\begin{figure*}
\centering
\includegraphics[width=\linewidth]{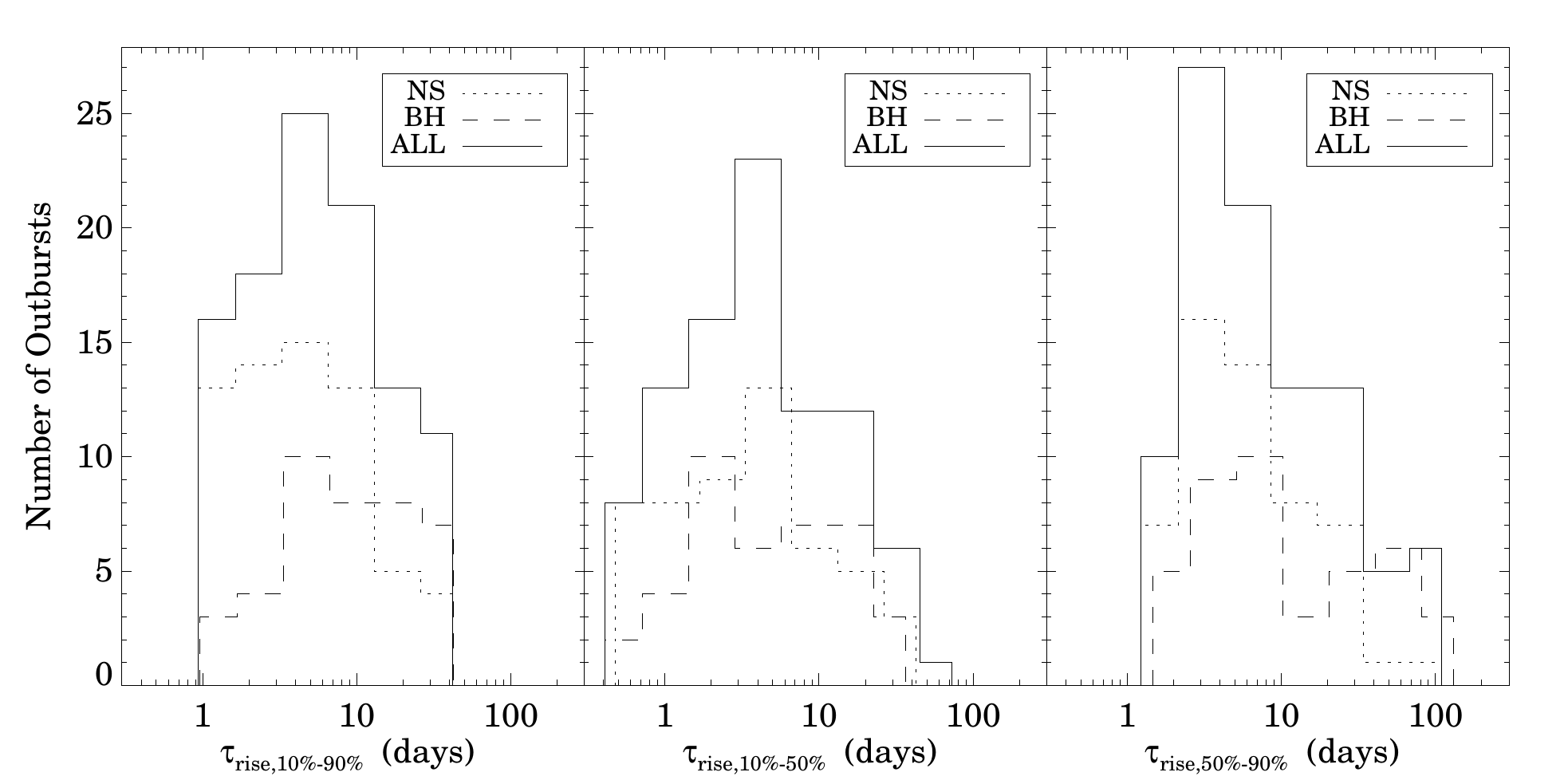}
\includegraphics[width=\linewidth]{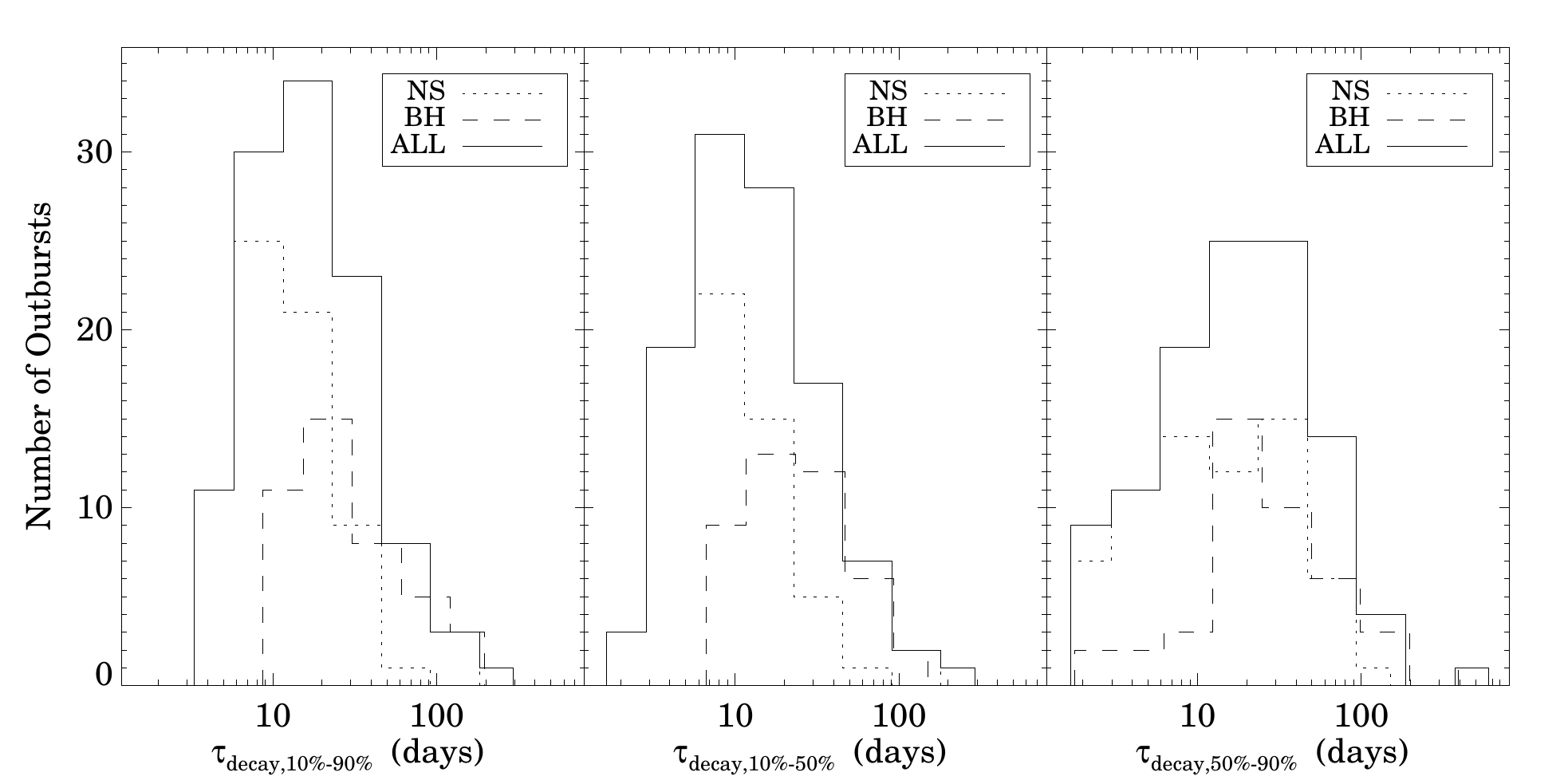}
\caption{The distributions of the $e$-folding rise or decay timescales during different rise or decay episodes. The averages of $\tau_\mathrm{rise,10\%-90\%}$, $\tau_\mathrm{rise,10\%-50\%}$, and $\tau_\mathrm{rise,50\%-90\%}$ in a logarithmic scale are 5.326, 3.672, and 7.038 days, respectively. The averages of $\tau_\mathrm{decay,10\%-90\%}$, $\tau_\mathrm{decay,10\%-50\%}$, and $\tau_\mathrm{decay,50\%-90\%}$ are 15.013, 12.194, and 15.330 days, respectively.}
\label{eft}
\end{figure*}

The $e$-folding time is calculated by
\begin{equation}
\label{eq_tau}
\tau = \frac{\Delta t}{\ln A}
\end{equation}
where $\Delta t$ is the time needed for the X-ray flux to increase by a factor of $A$. Note that the actual rise phase may not be well described by a single exponential form. We selected two data points in the light curve to measure the $e$-folding time for simplicity. Then $\tau$ corresponds to the $e$-folding time between the two data points in the light curve; that is, for an interval of $\tau$, the X-ray flux increases by a factor of $e\simeq2.71828$. The rise timescale defined in this way is independent of the instrumental sensitivity. 

In practice, from the outburst peak backward in time to the beginning of an outburst, we selected a data point marked as $F_{90\%}$, which has the closest flux to $90\%$ of the outburst peak flux $F_\mathrm{peak}$ in the range between the $F_\mathrm{peak}$ and the first two adjacent data points below 80\% of the $F_\mathrm{peak}$, so the  $F_{90\%}$ we selected could actually be in the range of $80\%-100\%~F_\mathrm{peak}$ in the real data. Then we selected a data point named $F_{50\%}$ which is the closest to $50\%$ of $F_\mathrm{peak}$ in the range between the first data point below 66\% of $F_\mathrm{peak}$ and the first two adjacent data points below 33\% of $F_\mathrm{peak}$, so the  $F_{50\%}$ could actually be in the range of $33\%-66\%~F_\mathrm{peak}$. We need to select two $F_{50\%}$ values if the outburst has a multipeak profile (e.g. the outburst of GRO J1655$-$40 in the \autoref{example1}).  A data point named $F_{10\%}$ with a flux closest to $10\%$ of $F_\mathrm{peak}$ was selected in the range between the first data point below 20\% of $F_\mathrm{peak}$ and the first two adjacent data points below 1\% of $F_\mathrm{peak}$, so the  $F_{10\%}$ could actually be in the range of $1\%-20\%~F_\mathrm{peak}$.  During the procedures indicated above, we excluded the data points in the daily light curves of unusual fluctuation for which a certain flux is larger or smaller than the fluxes of the two adjacent days by more than a factor of two (e.g., the data point on MJD 53589 of XTE J1818$-$245 and the data point on MJD 52859 of H1743$-$322 in the \autoref{example1}). Then the $e$-folding rise timescales $\tau_\mathrm{rise,10\%-90\%}$, $\tau_\mathrm{rise,10\%-50\%}$, and $\tau_\mathrm{rise,50\%-90\%}$ are calculated following \autoref{eq_tau}. 

In some outbursts, $F_{10\%}$ or $F_{50\%}$ cannot be identified because of data gaps (e.g., in \autoref{example1}, the data gaps during about MJD 50100--50130 in the outburst of GRS 1739$-278$ and the data gaps during about MJD 52600--52640 in the outburst of XTE J1720$-$318). There are other cases for which we could not identify the data points $F_{10\%}$ or $F_{50\%}$ because the X-ray flux increased too quickly on the daily timescale (e.g. the outburst of MXB 1730$-$333 in the \autoref{example1}). 

To reverse the order of the data during the decay phase, we used the same method as in the rise phase to select the $F_{90\%}$, $F_{50\%}$, and $F_{10\%}$. Then we obtained the $e$-folding decay timescales for the different decay episodes corresponding to $F_{10\%}$ -- $F_{90\%}$, $F_{10\%}$ -- $F_{50\%}$, and $F_{50\%}$ -- $F_{90\%}$, respectively. \autoref{eft} shows the distributions of the $e$-folding rise and decay timescales, including the BH LMXBTs, the NS LMXBTs, and all of the LMXBTs, respectively. \autoref{tau} lists the statistical results of the average $e$-folding rise and decay timescales in a logarithmic scale. Generally speaking, the $e$-folding decay timescale ($\sim$15 days) is about three times larger than the $e$-folding rise timescale ($\sim$5 days; see the $\tau_\mathrm{rise,10\%-90\%}$ and $\tau_\mathrm{decay,10\%-90\%}$ in the \autoref{tau}). From \autoref{eft} and \autoref{tau}, we can see that the $\tau_\mathrm{rise,10\%-50\%}$ is about two times smaller than $\tau_\mathrm{rise,50\%-90\%}$ (both for BH and NS LMXBTs). Both the $\tau_\mathrm{rise,10\%-90\%}$ and $\tau_\mathrm{rise,50\%-90\%}$ of BH LMXBTs are two times larger than those of NS LMXBTs, but the $\tau_\mathrm{rise,10\%-50\%}$ of BH and NS LMXBTs are similar. All the three $e$-folding decay timescales of BH LMXBTs are larger than those of NS LMXBTs by a factor of $\sim$ two. The $e$-folding decay timescales are similar between the decay episodes corresponding to $F_{10\%}$ -- $F_{50\%}$ and $F_{50\%}$ -- $F_{90\%}$ (both for BH and NS LMXBTs), which suggests that the decay phase (from $F_{90\%}$ to $F_{10\%}$ ) of most outbursts is roughly of the exponential form. 

The $e$-folding rise or decay timescale we measured might include the plateau phase or secondary flare for the outbursts with plateau or multipeak profiles. Some outbursts have a plateau or secondary flare during the decay (such as the outbursts of XTE J1701$-$462 and XTE J2012$+$381 in \autoref{example1}). It is quite interesting to study the decay before and after the plateau or secondary flare; this may reveal the mechanism of the plateau or the secondary flare. In the 1997 outburst of Aql X-1 and the 1998 outburst of XTE J2012$+$381, the $F_{90\%}$ -- $F_{50\%}$ and $F_{50\%}$ -- $F_{10\%}$ during the decay covered the phases before and after a short plateau or a secondary flare. The $\tau_\mathrm{decay,10\%-50\%}$ and $\tau_\mathrm{decay,50\%-90\%}$ were about 6 and 14 days for the outburst of Aql X-1, which means that the flux decreases faster after the plateau than before.  And $\tau_\mathrm{decay,10\%-50\%}$ and $\tau_\mathrm{decay,50\%-90\%}$ were about 18 and 19 days for the outburst of XTE J2012$+$381, which means that the flux decreases exponentially at the same rate before and after the secondary flare. \citet{chen97} has also mentioned that the source undergoes an exponential decay or sharp cutoff after a plateau phase.  Because a detailed study of the outburst profiles is beyond the of scope of this work, we cannot draw a conclusion about this issue at the moment.

\begin{figure*}
\centering
\includegraphics[width=\linewidth]{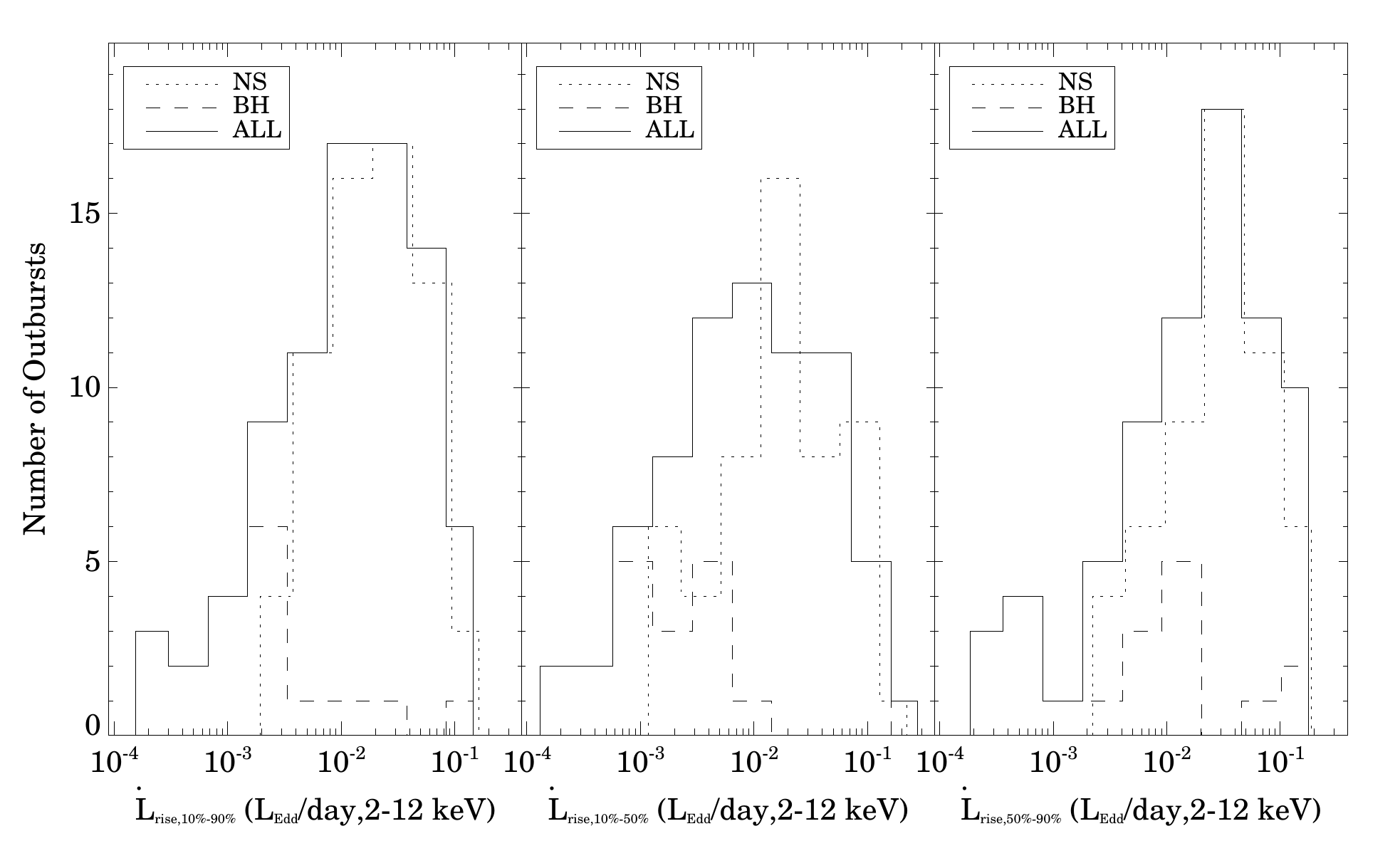}
\includegraphics[width=\linewidth]{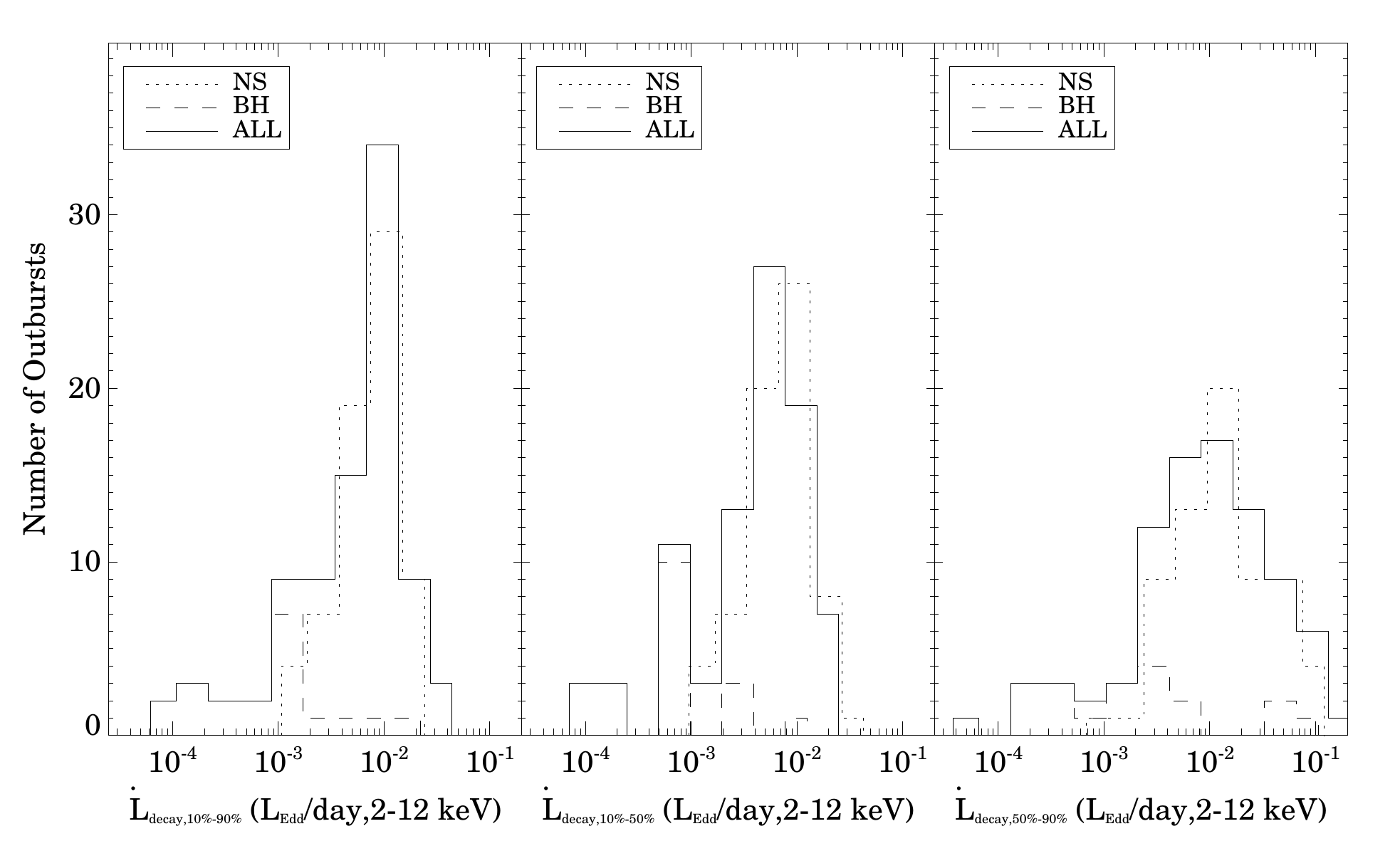}
\caption{The distributions of $\dot{L}$ in units of $L_\mathrm{Edd}$ during different rise or decay episodes. The averages of $\dot{L}_\mathrm{rise,10\%-90\%}$, $\dot{L}_\mathrm{rise,10\%-50\%}$, and $\dot{L}_\mathrm{rise,50\%-90\%}$ are 0.010, 0.008, and 0.015 $L_\mathrm{Edd}/\rm day$, respectively. The averages of $\dot{L}_\mathrm{decay,10\%-90\%}$, $\dot{L}_\mathrm{decay,10\%-50\%}$ and $\dot{L}_\mathrm{decay,50\%-90\%}$ are 0.004, 0.003, and 0.008 $L_\mathrm{Edd}/\rm day$, respectively.}
\label{t_ldot}
\end{figure*}

We also measured the rate of change  of flux according to $\dot{F}=\Delta F/\Delta t$ for the different rise or decay episodes corresponding to $F_{10\%}$ -- $F_{50\%}$, $F_{50\%}$ -- $F_{90\%}$, and $F_{10\%}$ -- $F_{90\%}$, respectively. For the sources with known distances and masses, we scaled the $\dot{L}$ in units of $L_\mathrm{Edd}$. \autoref{t_ldot} shows the distributions of $\dot{L}$ of different rise and decay episodes in a logarithmic scale, and \autoref{ldot} shows the statistical results of the average $\dot{L}$ in a logarithmic scale. From \autoref{t_ldot} and \autoref{ldot}, we can see that the $\dot{L}_\mathrm{rise,50\%-90\%}$ is about two times larger than $\dot{L}_\mathrm{rise,10\%-50\%}$ for both the BH and NS LMXBTs, and $\dot{L}_\mathrm{decay,50\%-90\%}$ is also about two times larger than $\dot{L}_\mathrm{decay,10\%-50\%}$. These suggest that the X-ray luminosity changes much more dramatically when it is near the peak X-ray luminosity.

\begin{figure*}
\centering
\plottwo{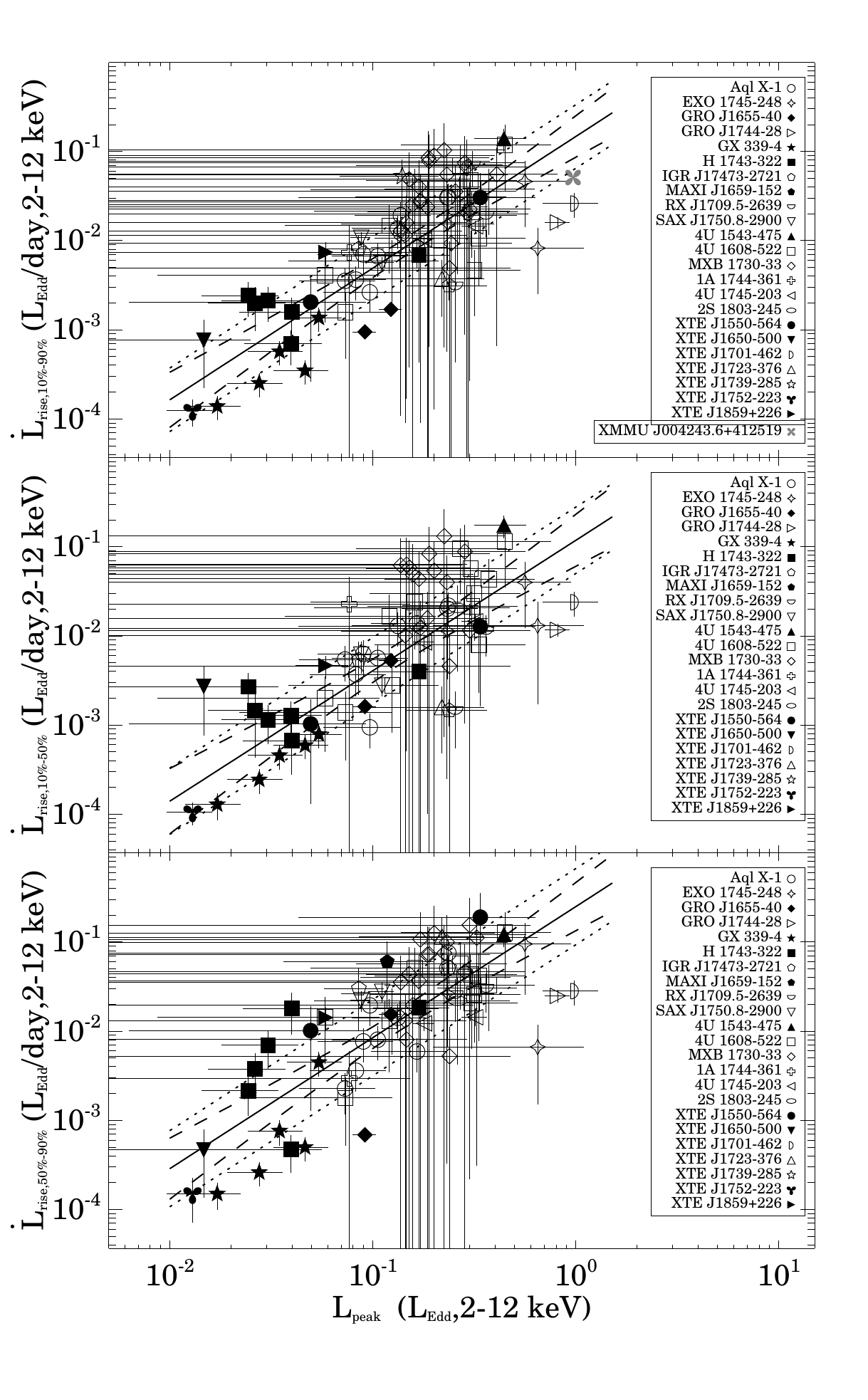}{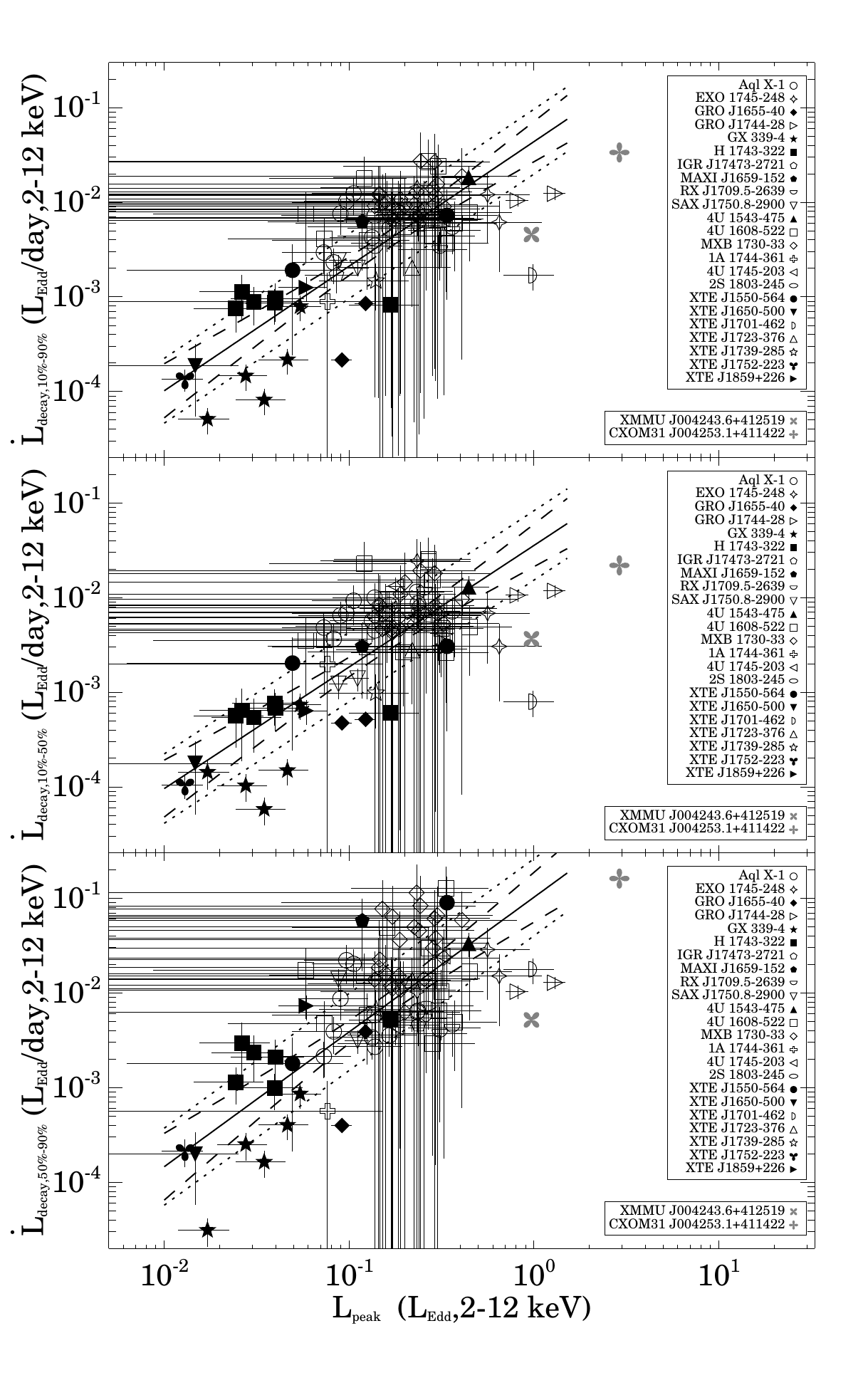}
\caption{Relation between $\dot{L}$ and $L_\mathrm{peak}$ in units of $L_\mathrm{Edd}$. The filled and unfilled symbols represent BH and NS LMXBTs, respectively. There are positive correlations between $L_\mathrm{peak}$ and $\dot{L}$ in different rise or decay episodes. The solid line represents the best-fit result with a function $\log \dot{L} = A+B\times\log L_\mathrm{peak}$, the dashed lines show the 2$\sigma$ confidence intervals, and the dotted lines show the range of plus or minus intrinsic scatter}
\label{ldot_peak}
\end{figure*}

\begin{figure}
\centering
\includegraphics[width=\linewidth]{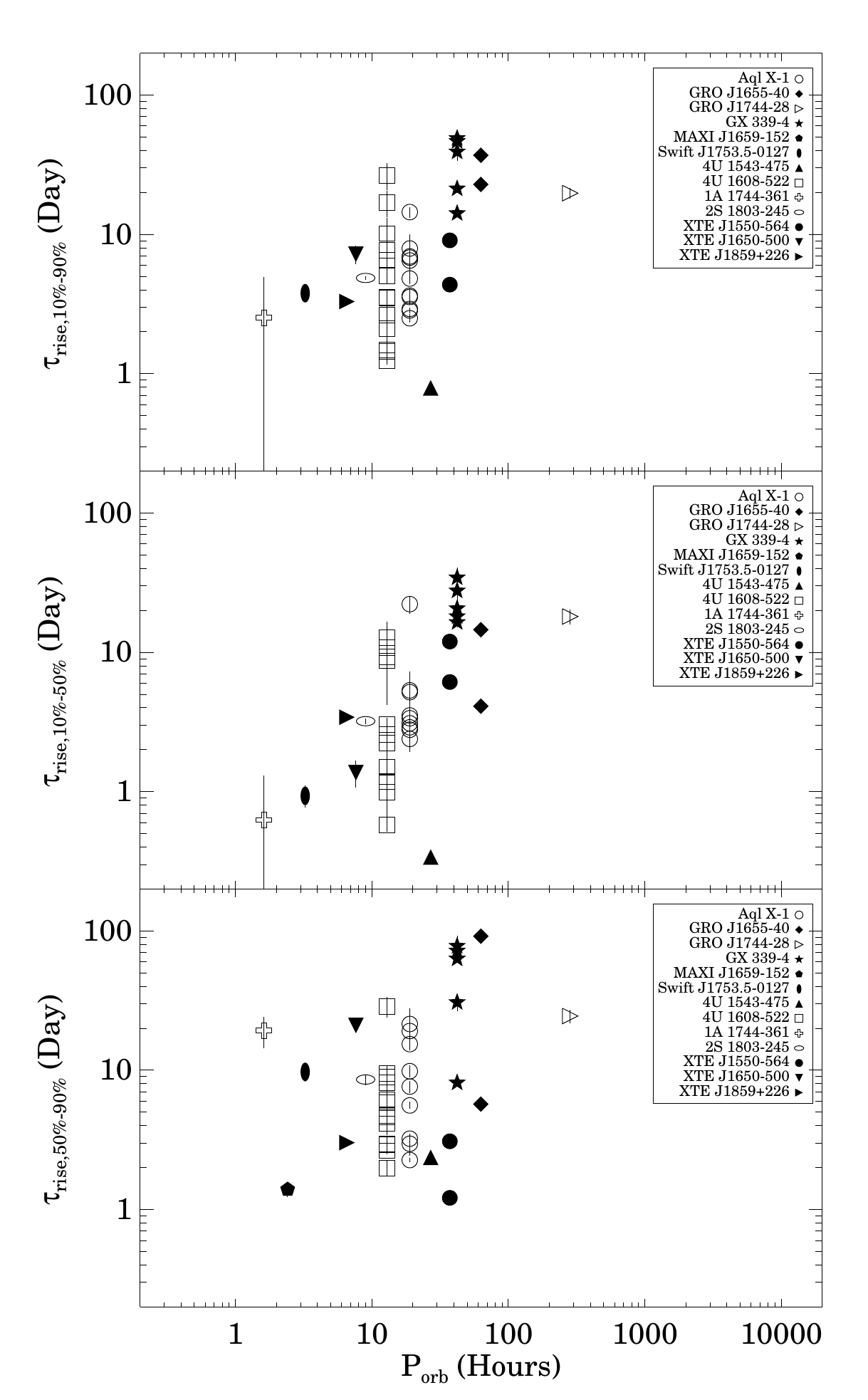}
\caption{Relation between orbital period and $e$-folding rise timescale during different rise episodes. The filled and unfilled
symbols represent BH and NS LMXBTs, respectively. There are weak correlations between $P_\mathrm{orb}$ and $\tau_\mathrm{rise}$ in different rise episodes.}
\label{orb_rise}
\end{figure}

We showed the correlation between $\dot{L}_\mathrm{rise}$ and $L_\mathrm{peak}$ scaled in $L_\mathrm{Edd}$ in \autoref{ldot_peak}. The Spearman correlation coefficients of these three correlations are $0.70$, $0.67$, and $0.71$ at a significance of $7.28 \sigma$, $6.31 \sigma$, and $6.87 \sigma$ for the different rise episodes corresponding to $F_{10\%}$ -- $F_{90\%}$, $F_{10\%}$ -- $F_{50\%}$, and $F_{50\%}$ -- $F_{90\%}$, respectively. These results demonstrate that there is a significant positive correlation between $\dot{L}_\mathrm{rise}$ and $L_\mathrm{peak}$ (see \autoref{ldot_peak}), which confirms the previous results in \citet{yy09}. We used a linear function in a logarithmic scale ($\log \dot{L}_\mathrm{rise} = A+B\times\log L_\mathrm{peak}$)  to fit the three correlations with a Bayesian approach following \citet{kelly07}.  Then we got $\log \dot{L}_\mathrm{rise,10\%-90\%} =( -0.83\pm0.14)+(1.48\pm0.15)\times\log L_\mathrm{peak}$, $\log \dot{L}_\mathrm{rise,10\%-50\%} =( -0.93\pm0.15)+(1.47\pm0.16)\times\log L_\mathrm{peak}$, and $\log \dot{L}_\mathrm{rise,50\%-90\%} =( -0.59\pm0.17)+(1.48\pm0.17)\times\log L_\mathrm{peak}$, respectively. The intrinsic scatters of these correlations  are 0.36$\pm$0.05, 0.37$\pm$0.06, and 0.43$\pm$0.06 dex, respectively. The best-fit results are also plotted in \autoref{ldot_peak}. We found that the $L_\mathrm{peak}$ is also positively correlated with $\dot{L}_\mathrm{decay}$ in units of $L_\mathrm{Edd}$ (see \autoref{ldot_peak}). The Spearman correlation coefficients are $0.59$, $0.56$, and $0.60$ at a significance of $5.84\sigma$, $5.39 \sigma$, and $5.97 \sigma$ for the different decay episodes corresponding to $F_{10\%}$ -- $F_{90\%}$, $F_{10\%}$ -- $F_{50\%}$, and $F_{50\%}$ -- $F_{90\%}$, respectively, which shows that the positive correlations between $L_\mathrm{peak}$ and $\dot{L}_\mathrm{decay}$ are also very significant.  We then used the same function ($\log \dot{L}_\mathrm{decay} = A+B\times\log L_\mathrm{peak}$)  to fit the three correlations with a Bayesian approach in \citet{kelly07}, and we got $\log \dot{L}_\mathrm{decay,10\%-90\%} =( -1.36\pm0.14)+(1.32\pm0.15)\times\log L_\mathrm{peak}$, $\log \dot{L}_\mathrm{decay,10\%-50\%} =( -1.44\pm0.15)+(1.29\pm0.16)\times\log L_\mathrm{peak}$, and $\log \dot{L}_\mathrm{decay,50\%-90\%} =( -0.98\pm0.16)+(1.43\pm0.17)\times\log L_\mathrm{peak}$, respectively. The intrinsic scatters of the three correlations are 0.34$\pm$0.04, 0.36$\pm$0.05, and 0.40$\pm$0.05 dex, respectively. The best-fit results are also plotted in \autoref{ldot_peak}. In \autoref{ldot_peak}, it seems that these positive correlations are also hold in the individual source GX 339$-$4. In addition, we also found a positive correlation between the $e$-folding rise timescale and the orbital period (see \autoref{orb_rise}). The Spearman correlation coefficients are 0.51, 0.63, and 0.29 at a significance of $3.15 \sigma$, $3.99 \sigma$, and $1.45 \sigma$ for $\tau_{10\%-90\%}$, $\tau_{10\%-50\%}$, and $\tau_{50\%-90\%}$, respectively, but no correlation between the $e$-folding decay timescale and the orbital period was found.

\capstartfalse
\begin{deluxetable*}{lccc}
\centering
\tablecolumns{5}
\tabletypesize{\footnotesize}
\tablecaption{Statistical Results of the Average $\dot{L}$ }
\tablewidth{0pt}
\tablehead{
\colhead { } & \colhead{All LMXBTs}  & \colhead{NS LMXBTs} &\colhead{BH LMXBTs}  
}

\startdata
$<\log(\dot{L}_\mathrm{rise,10\%-90\%})>$ & $-1.994\pm 0.692$ & $-1.756\pm 0.457 $ & $-2.797\pm 0.755$ \\
$<\dot{L}_\mathrm{rise,10\%-90\%}> (L_\mathrm{Edd}/\rm day)$ & $ 0.010^{+ 0.040}_{- 0.008}$ & $0.018^{+ 0.033}_{- 0.011}$ &$0.002^{+ 0.007}_{- 0.001}$ \\
$<\log(\dot{L}_\mathrm{rise,10\%-50\%})>$ & $-2.108\pm 0.742$ & $-1.848\pm 0.553$ & $-2.820\pm 0.736$ \\
$<\dot{L}_\mathrm{rise,10\%-50\%}> (L_\mathrm{Edd}/\rm day)$ & $0.008^{+ 0.035}_{- 0.006}$ & $ 0.014^{+ 0.037}_{- 0.010}$ & $0.002^{+ 0.007}_{- 0.001}$ \\
$<\log(\dot{L}_\mathrm{rise,50\%-90\%})>$ & $-1.819\pm 0.740$ & $-1.584\pm 0.468$ & $-2.453\pm 0.954$ \\
$<\dot{L}_\mathrm{rise,50\%-90\%}> (L_\mathrm{Edd}/\rm day)$ & $0.015^{+ 0.068}_{- 0.012}$ & $ 0.026^{+ 0.051}_{- 0.017}$ &$ 0.004^{+ 0.028}_{- 0.003}$ \\
\hline
$<\log(\dot{L}_\mathrm{decay,10\%-90\%})>$ & $-2.379\pm 0.595$ & $-2.148\pm 0.313 $ & $-3.164\pm 0.660$ \\
$<\dot{L}_\mathrm{decay,10\%-90\%}> (L_\mathrm{Edd}/\rm day)$ & $0.004^{+ 0.012}_{- 0.003}$ & $ 0.007^{+ 0.008}_{- 0.004}$ &$0.001^{+ 0.002}_{- 0.001}$ \\
$<\log(\dot{L}_\mathrm{decay,10\%-50\%})>$ & $-2.458\pm 0.594$ & $-2.212\pm 0.316$ & $-3.269\pm 0.575$ \\
$<\dot{L}_\mathrm{decay,10\%-50\%}> (L_\mathrm{Edd}/\rm day)$ & $0.003^{+ 0.010}_{- 0.003}$ & $0.006^{+ 0.007}_{- 0.003}$ & $0.001^{+ 0.001}_{-0.0004}$ \\
$<\log(\dot{L}_\mathrm{decay,50\%-90\%})>$ & $-2.102\pm 0.707$ & $-1.887\pm 0.466$ & $-2.813\pm 0.894$ \\
$<\dot{L}_\mathrm{decay,50\%-90\%}> (L_\mathrm{Edd}/\rm day)$ & $0.008^{+ 0.032}_{- 0.006}$ & $ 0.013^{+ 0.025}_{- 0.009}$ &$ 0.002^{+ 0.011}_{- 0.001}$ \\
\enddata
\label{ldot}
\end{deluxetable*}
 \capstarttrue
 
\subsection{Outburst Duration}

\capstartfalse
\begin{deluxetable*}{lccc}
\centering
\tablecolumns{5}
\tabletypesize{\footnotesize}
\tablecaption{Statistical Results of the Average Outburst Duration }
\tablewidth{0pt}
\tablehead{
\colhead { } & \colhead{All LMXBTs}  & \colhead{NS LMXBTs} &\colhead{BH LMXBTs}  
}

\startdata
$<\log(\rm Duration)>$ & $1.729\pm 0.355$ & $1.595\pm 0.307 $ & $1.945\pm 0.321$ \\
$<\rm Duration>$ (days) & $53.641^{+67.904}_{-29.968}$ & $39.316^{+40.436}_{-19.934}$ &$88.183^{+96.579}_{-46.095}$ \\

\enddata
\label{t_dur}
\end{deluxetable*}
 \capstarttrue

\begin{figure}
\centering
\includegraphics[width=\linewidth]{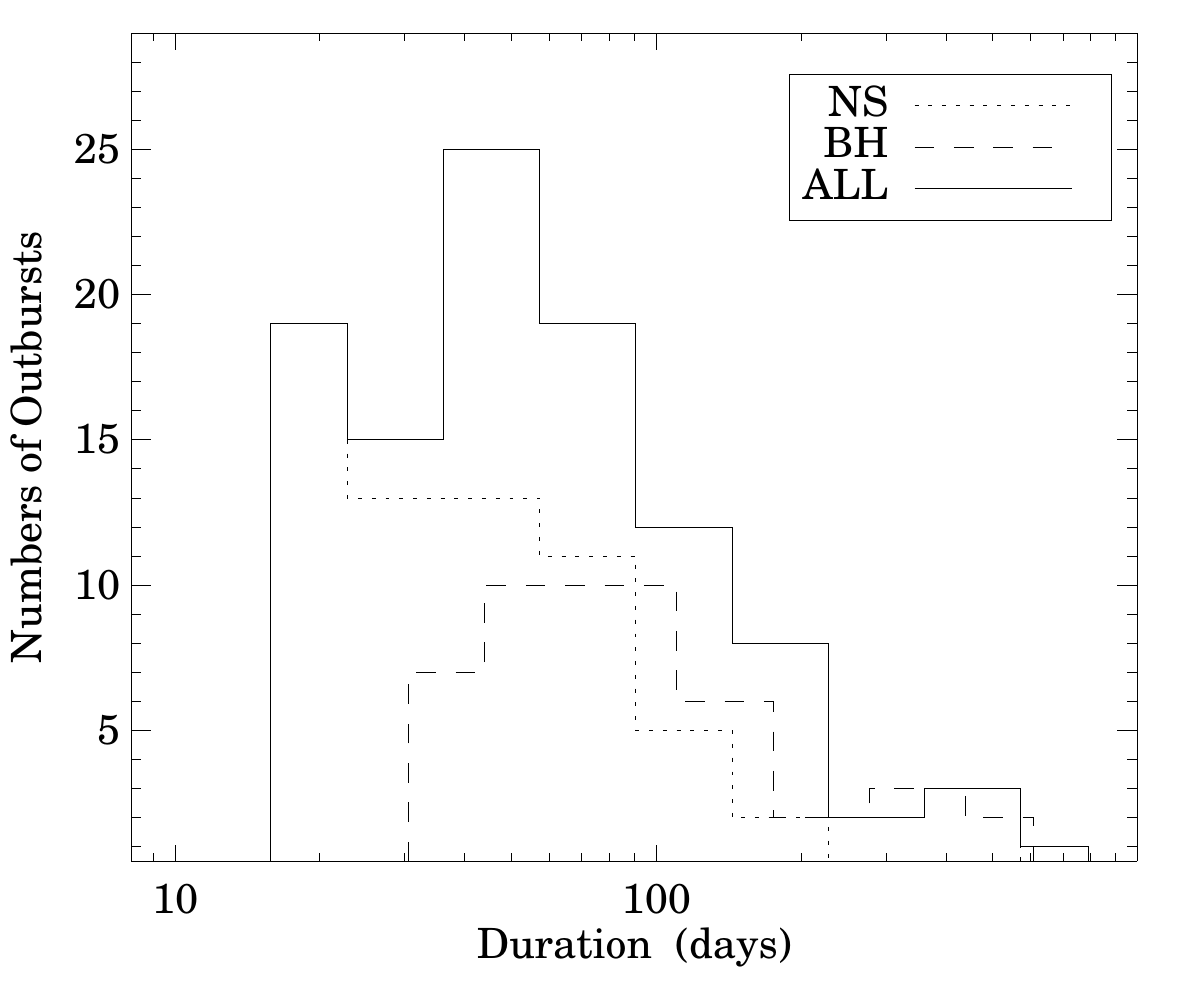}
\caption{The distribution of the outburst duration. The average of outburst duration is about 53.641, 39.316 and 88.183 days for all of the LMXBTs, NSs and BHs respectively.}
\label{d_dur}
\end{figure}

\begin{figure}
\centering
\includegraphics[width=\linewidth]{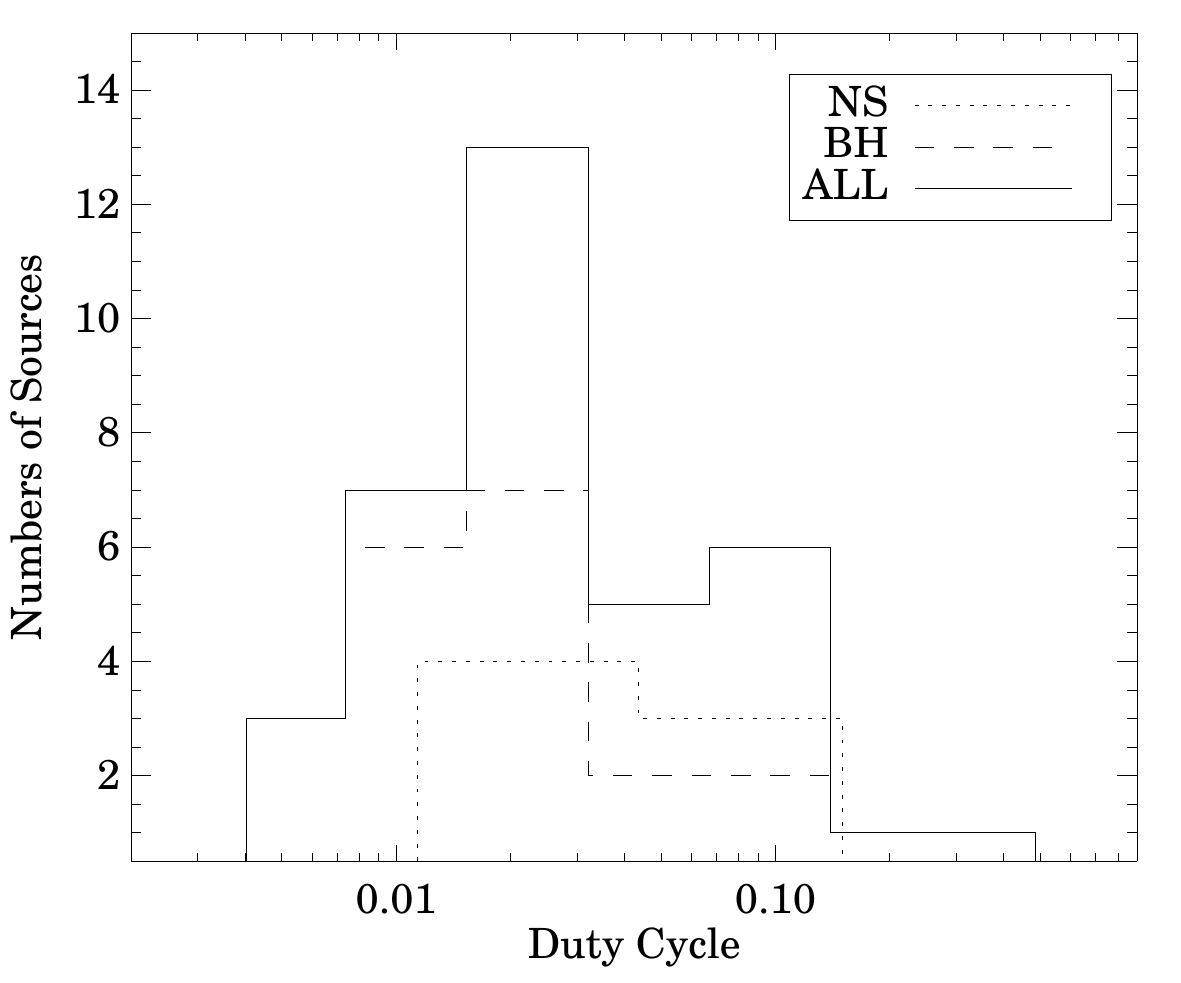}
\caption{The distribution of the duty cycles. The average duty cycle is about 0.025, 0.035 and 0.020 for all of the LMXBTs, NSs and BHs respectively.}
\label{dc}
\end{figure}

For outbursts with $F_{10\%}$ identified both in the rise and the decay phase, we estimated the duration of an outburst as the time interval between these two $F_{10\%}$ values. In this way, our estimation of outburst duration is independent of the instrument sensitivity. \autoref{d_dur} shows the distributions of the outburst durations for the available outbursts, and \autoref{t_dur} shows the statistical results of the average duration in a logarithmic scale. Based on our definition, the average duration of outbursts in LMXBTs is about 54 days. From \autoref{d_dur} and \autoref{t_dur}, we can see that the average duration of BH LMXBTs ($\sim$88 days) is larger than that of NS LMXBTs ($\sim$39 days) by a factor of about two. 

We estimated the duty cycle for each LMXBT using the outburst duration and the number of outbursts we measured. This parameter is very important in understanding the luminosity function and evolution of LMXBs \citep[such as ][]{Belczynski2004,Fragos2008,Fragos2009}, which is defined as the fraction of the lifetime that a transient source is in the outburst phase. Because of the highly reduced sky coverage of {\it RXTE}/ASM since late 2010, we only used the data before 2011 to estimate the duty cycle. We summed all of the outburst durations during the period 1996 -- 2010 and divided by the total observation time ($\sim$ 5472 days) to estimate the duty cycle for each LMXBT. For the outbursts, we could not measure the outburst duration because of the lack of identification of $F_{10\%}$ in the rise or decay phases, so we used the $F_{50\%}$ or $F_{90\%}$ (if lacking $F_{50\%}$) instead. 

\autoref{dc} shows the distribution of the duty cycles we estimated for all of the 36 LMXBTs. The average of duty cycles in a logarithmic scale is $0.025^{+ 0.045}_{- 0.016}$, $0.035^{+ 0.044}_{- 0.019}$, and $0.020^{+ 0.042}_{- 0.014}$ for all LMXBTs, NSs, and BHs, respectively. However, many LMXBTs in our sample only have one outburst during the entire $RXTE$ era, so the duty cycle we estimated is actually the upper limit of the true value. The evolution history of the LMXBTs is much longer than the 15 yr long $RXTE$/ASM observation, so the behavior during the $RXTE$ era may not be typical in the long evolution history. We do at least get an order-of-magnitude estimate of the duty cycles, which is helpful in the theoretical modeling of the luminosity function and the evolution of LMXBTs. 

Some sources in our sample were also investigated by \citet{chen97}, including two NS LMXBTs, Aql X-1 and 4U 1608$-$522, and three BH LMXBTs, 4U 1543$-$475, 4U 1630$-$472, and GRO J1655$-$40.  We have combined the results in \citet{chen97} to improve the estimation of the duty cycles of these sources.  We used the average outburst duration in the $RXTE$ era for outbursts with no duration measurements in \citet{chen97}. The duty cycles of Aql X-1 and 4U 1608$-$522 from 1970 to 2011 are about 0.08 and 0.09, which are the same as that in the $RXTE$ era. The duty cycle of 4U 1543$-$475 from 1971 to 2011 is about 0.05, which is one order of magnitude larger than that in the $RXTE$ era. The duty cycles of GRO J1655$-$40 from 1994 to 2011 and 4U 1630$-$472 from 1971 to 2011 are about 0.17 and 0.21, which are larger than the 0.12 and 0.16 estimated for the outbursts only in the $RXTE$ era. 

\citet{Grindlay2014} has discovered historical optical outbursts in four BH LMXBTs. The data of XTE J1118$+$480 are released in the DASCH (Digital Access to a Sky Century @ Harvard) project.  We identified 10 optical outbursts during the period from about 1911 to 1989 from the DASCH data of XTE J1118$+$480. There is an X-ray outburst of XTE J1118$+$480 during the $RXTE$ era, which is not included in our sample because the peak flux was less than 0.1 Crab. If we assume that all of the historical optical outbursts were associated with X-ray outbursts and the outburst durations of those optical outbursts are similar to that of the X-ray outbursts in the $RXTE$ era, the duty cycle of XTE J1118$+$480 from 1911 to 2011 is about 0.05, which is a little larger than the estimation (0.03)  from only the data of the $RXTE$ era.

\subsection{Total Energy Radiated during Outbursts }
We integrated the X-ray flux over an outburst from $F_{10\%}$ in the rise phase to $F_{10\%}$ in the decay phase to estimate the total X-ray fluence for each available outburst. The X-ray energy spectra of LMXBTs usually varies dramatically during an outburst. In order to estimate the X-ray flux more accurately, we extracted the ASM light curves in three energy bands and scaled the X-ray flux in Crab units with conversion factors of 1 Crab=27, 23, and 25 c s$^{-1}$ for 2--3 keV, 3--5 keV, and 5--12 keV, respectively. Then we converted the photon fluxes of three energy bands into fluxes by assuming a Crab Nebula-like X-ray spectrum, and we summed the fluxes in the three energy bands to obtain the X-ray flux in the 2--12 keV band. Then we measured the total X-ray fluence by summing the 2--12 keV X-ray flux from $F_{10\%}$ in the rise phase to $F_{10\%}$ in the decay phase. The data gaps in the daily light curve due to sparse coverage were filled with values inferred by linear interpolation. For sources with known distances, we obtained the total energy $E$ radiated during the outburst. We also measured the total radiated energy during the rise and decay phases in this way. 

\begin{figure*}
\centering
\includegraphics[width=\linewidth]{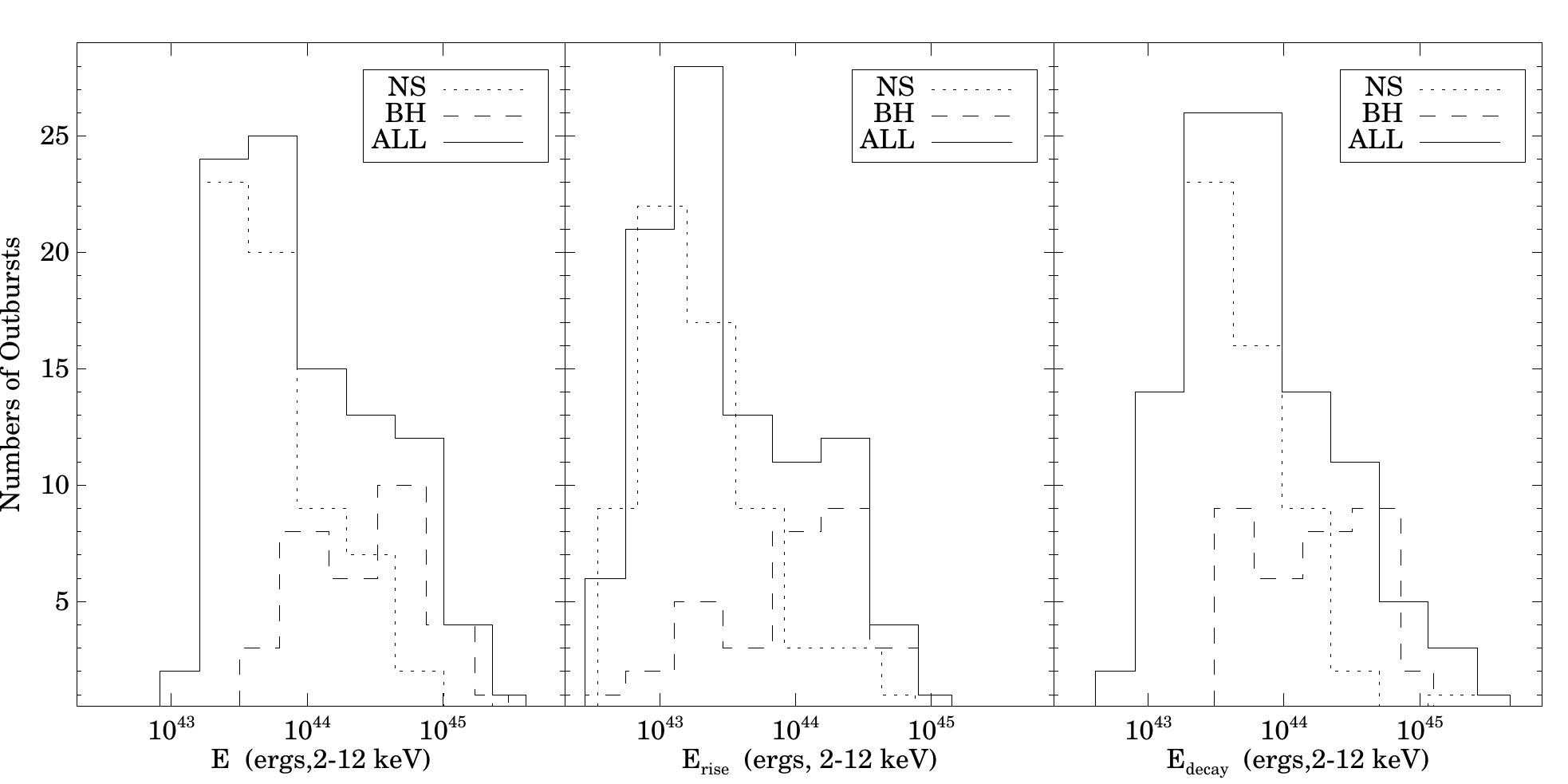}
\caption{The distributions of the total radiated energy in 2--12 keV. The averages of $E$ are $0.908\times10^{44}$ ergs, $0.551\times10^{44}$ ergs, and $2.468\times10^{44}$ ergs for all of the LMXBTs, NSs, and BHs, respectively.}
\label{h_te}
\end{figure*}

\autoref{h_te} shows the distributions of the total radiated energy $E$ in the 2--12 keV band. \autoref{t_te} shows the statistical results of the average total radiated energy in a logarithmic scale. As can be seen from \autoref{t_te} and \autoref{h_te}, the average total energy of BH LMXBTs in 2--12 keV ($\sim2.47\times10^{44}$ ergs) is about five times larger than that of NS LMXBTs ($\sim0.55\times10^{44}$ ergs). The average total energy is about $0.91\times10^{44}$ ergs. Assuming a bolometric X-ray flux correction factor of three and a constant radiative efficiency of 0.1, this corresponds to a total mass of $1.52\times10^{-9} M_{\sun}$ on average being accreted during an outburst. The total radiated energy during the decay phase is about two times larger than that during the rise phase for both BH and NS LMXBTs, which suggests that the central compact object will accrete more mass during the decay phase (see \autoref{t_te}). There is a cutoff of the distribution of $E$ for BH LMXBTs at the lower $E$ end, but it does not show in the NS LMXBT distribution (see left panel of \autoref{h_te}), which may suggest that there is a minimum disk mass is required to trigger an outburst for BH LMXBTs.
\capstartfalse
\begin{deluxetable*}{lccc}
\centering
\tablecolumns{5}
\tabletypesize{\footnotesize}
\tablecaption{Statistical Results of the Average Total Radiated Energy}
\tablewidth{0pt}
\tablehead{
\colhead { } & \colhead{All LMXBTs}  & \colhead{NS LMXBTs} &\colhead{BH LMXBTs}  
}

\startdata
$<\log(E_{total})>$ & $43.958\pm 0.565$ & $43.741\pm 0.468$ & $44.392\pm 0.491$ \\
$<E_{total}> (10^{44} ergs,2-12 keV)$ & $ 0.908^{+ 2.425}_{- 0.660}$ & $0.551^{+ 1.065}_{- 0.363}$ &$2.468^{+ 5.180}_{- 1.671}$ \\
$<\log(E_\mathrm{rise})>$ & $43.469\pm 0.594$ & $43.253\pm 0.476$ & $43.901\pm 0.577$ \\
$<E_\mathrm{rise}> (10^{44} ergs,2-12 keV)$ & $0.295^{+ 0.862}_{- 0.220}$ & $0.179^{+ 0.357}_{- 0.119}$ &$0.796^{+ 2.206}_{- 0.585}$ \\
$<\log(E_\mathrm{decay})>$ & $43.795\pm 0.580$ & $43.597\pm 0.520$ & $44.191\pm 0.489$ \\
$<E_\mathrm{decay}> (10^{44} ergs,2-12 keV)$ & $0.624^{+ 1.749}_{- 0.460}$ & $0.396^{+ 0.913}_{- 0.276}$ &$1.554^{+ 3.236}_{- 1.050}$ \\
\enddata
\label{t_te}
\end{deluxetable*}
\capstarttrue
 
\begin{figure}
\centering
\includegraphics[width=\linewidth]{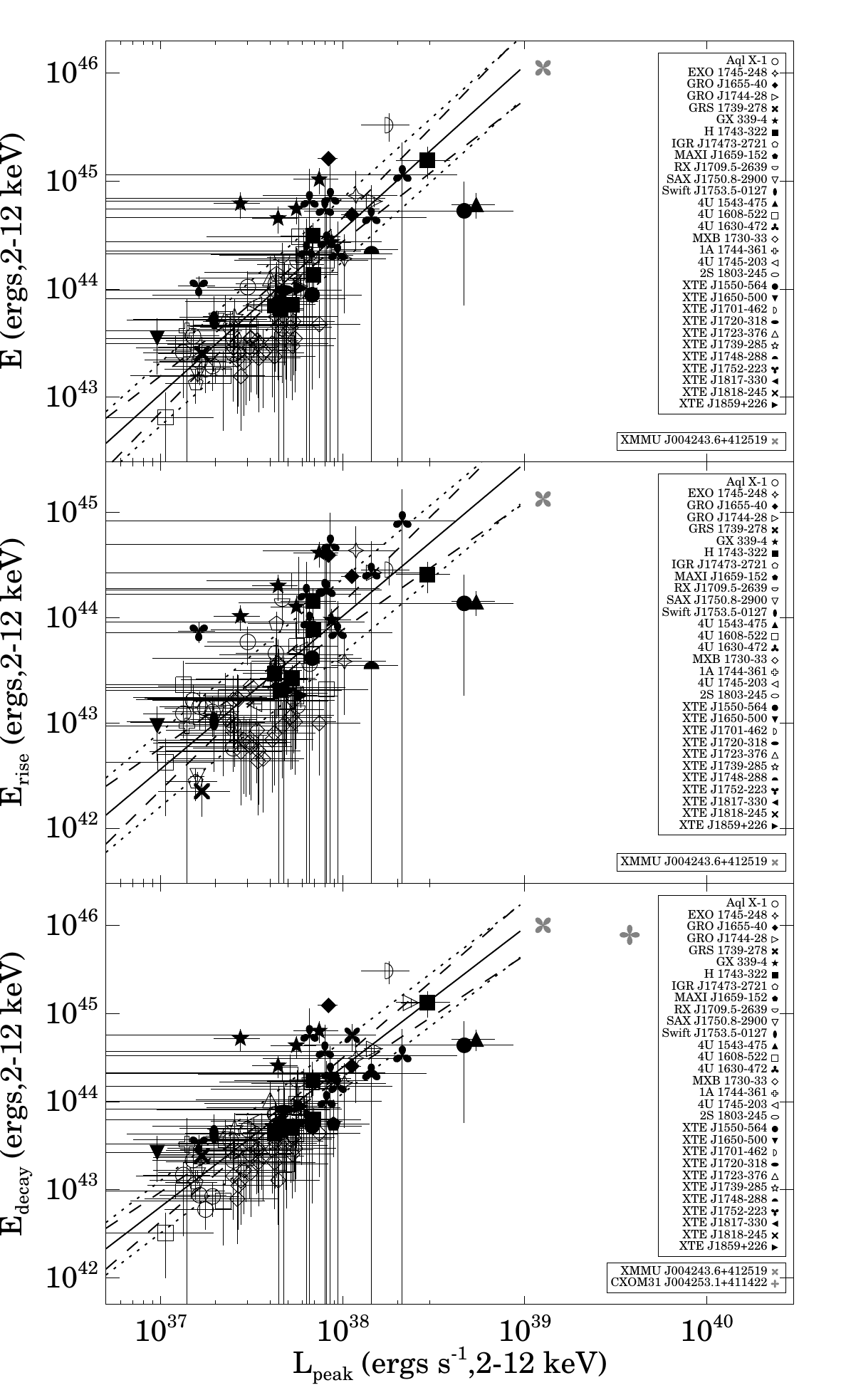}
\caption{Relation between peak X-ray luminosity and total radiated energy. The filled and unfilled
symbols represent BH and NS LMXBTs, respectively. There are positive correlations between the peak X-ray luminosity and the total radiated energy. The solid line represents the best-fit result with a function $\log E = A+B\times\log L_\mathrm{peak}$, the dashed lines show the 2$\sigma$ confidence intervals, and the dotted lines show the range of plus or minus intrinsic scatter.} 
\label{p_te}
\end{figure}

\begin{figure*}
\centering
\plottwo{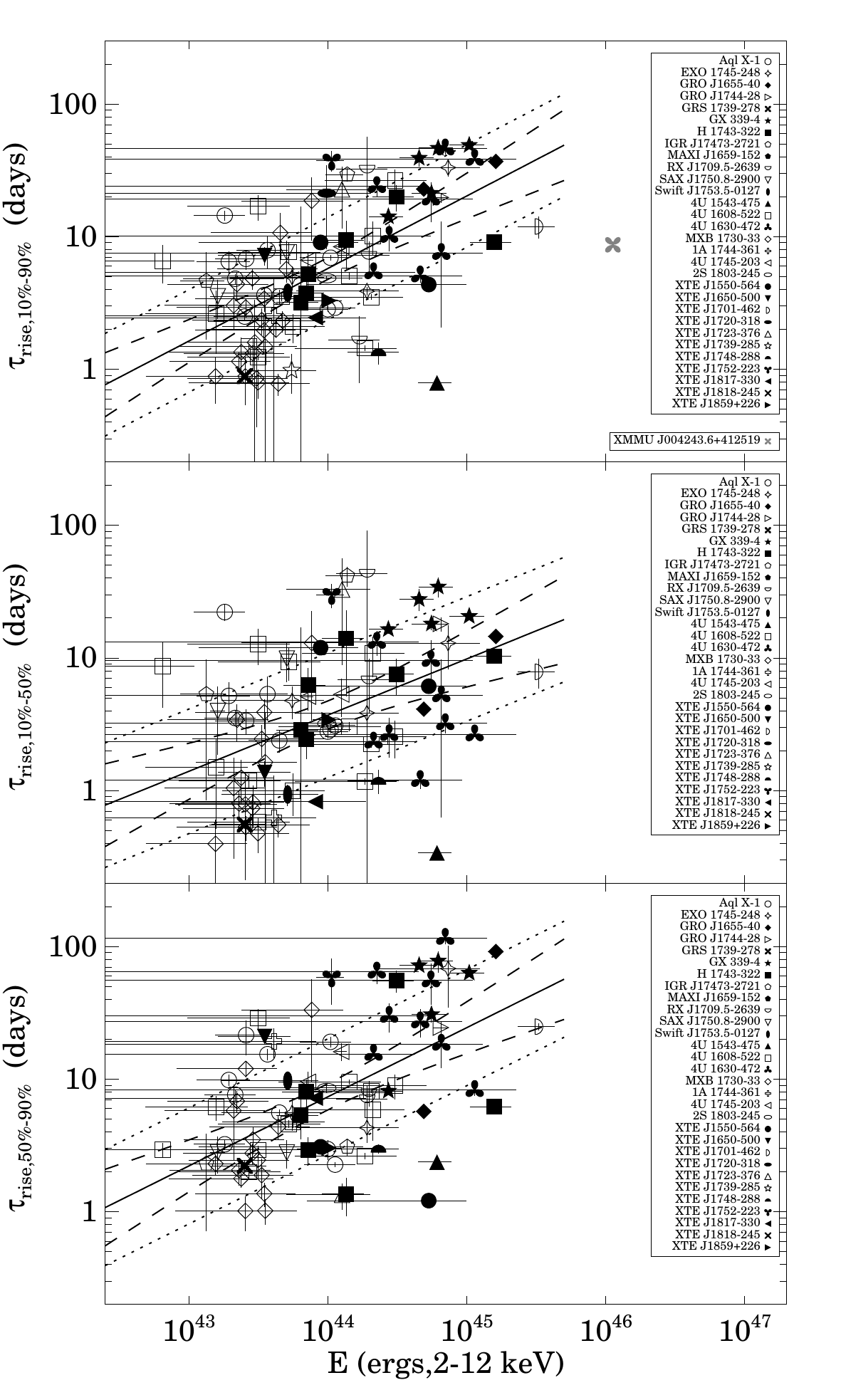}{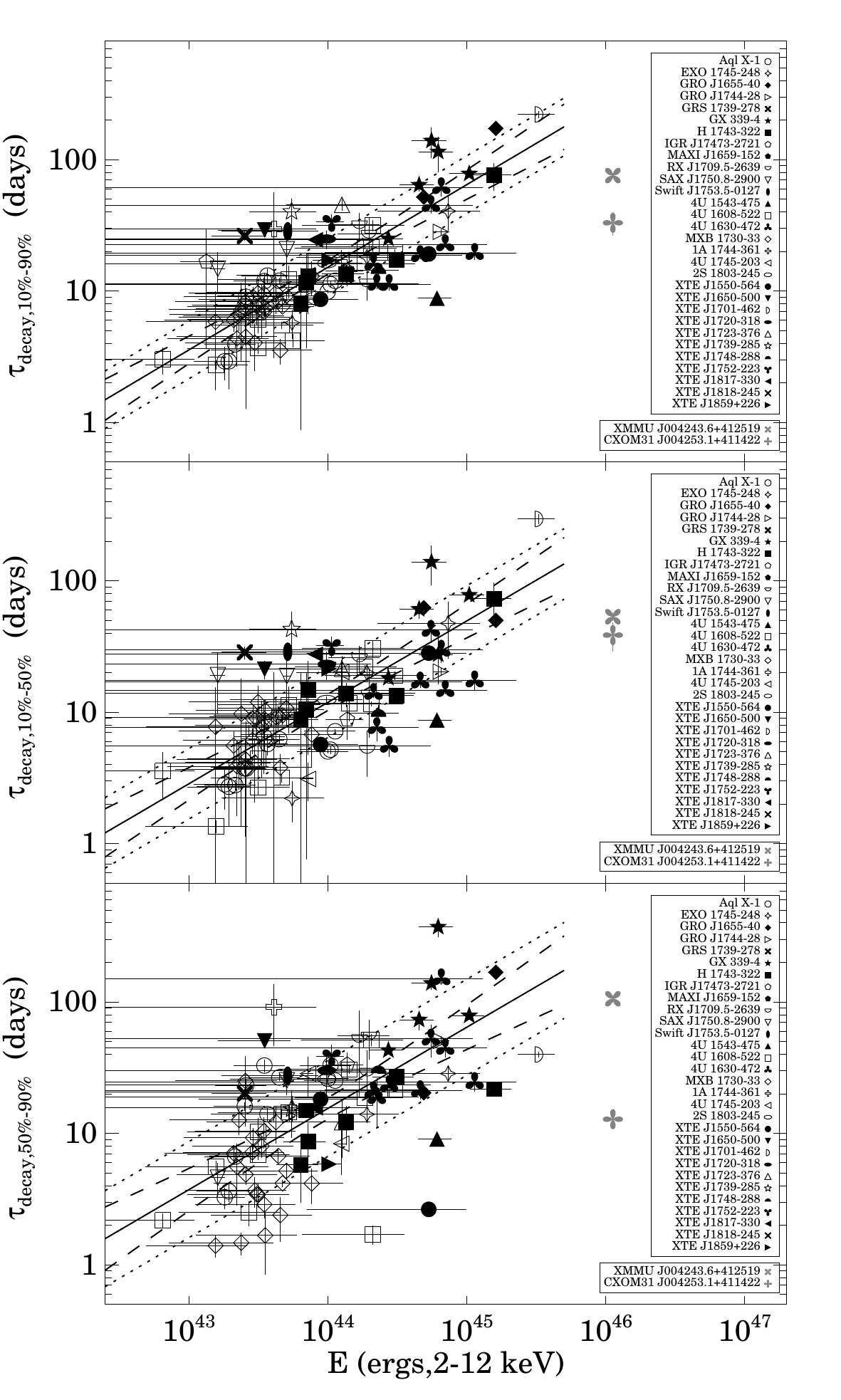}
\caption{Relation between $E$-folding rise or timescale and total radiated energy. The filled and unfilled
symbols represent BH and NS LMXBTs, respectively. There are positives correlation between $E$ and $\tau$ in different rise or decay episodes.The solid line represents the best-fit result with a function $\log \tau = A+B\times\log E$, the dashed lines show the 2$\sigma$ confidence intervals, and the dotted lines show the range of plus or minus intrinsic scatter.}
\label{tau_te}
\end{figure*}

We further plotted the correlation between the peak X-ray luminosity and total radiated energy (see \autoref{p_te}). The Spearman correlation coefficients are $0.79$, $0.71$, and $0.80$ at a significance of $9.51 \sigma$, $8.00 \sigma$, and $10.02 \sigma$ for the $E$, $E_\mathrm{rise}$, and $E_\mathrm{decay}$, respectively, which demonstrate that there is a strong positive correlation between the peak X-ray luminosity and the total radiated energy. We used a linear function in a logarithmic scale ($\log E = A+B\times\log L_\mathrm{peak}$)  to fit the three correlations with a Bayesian approach in \citet{kelly07}.  Then we got $\log E =(-13.26\pm 6.53)+(1.52\pm0.17)\times\log L_\mathrm{peak}$, $\log E_\mathrm{rise} =(-11.18\pm7.31)+(1.45\pm0.19)\times\log L_\mathrm{peak}$ , and $\log E_\mathrm{decay} =(-15.76\pm6.40)+(1.58\pm0.17)\times\log L_\mathrm{peak}$, respectively. The intrinsic scatters of these correlations are 0.29$\pm$0.04, 0.35$\pm$0.05, and 0.30$\pm$0.04 dex, respectively. The best-fit results are also plotted in \autoref{p_te}. These correlations seem to be also hold in an individual source MXB 1730-33 (see \autoref{p_te}). We also found that the total radiated energy is positively correlated with the $e$-folding rise and decay timescale (see \autoref{tau_te}). For the rise timescale, the Spearman correlation coefficients are $0.57$, $0.42$, and $0.49$ at a significance of $5.97 \sigma$, $3.78 \sigma$, and $4.55 \sigma$ for $\tau_\mathrm{rise,10\%-90\%}$, $\tau_\mathrm{rise,10\%-50\%}$, and $\tau_\mathrm{rise,50\%-90\%}$, respectively. For the decay timescale, the Spearman correlation coefficients are $0.73$, $0.66$, and $0.63$ at a significance of $8.43 \sigma$, $7.12 \sigma$, and $6.69 \sigma$ for $\tau_\mathrm{decay,10\%-90\%}$, $\tau_\mathrm{decay,10\%-50\%}$, and $\tau_\mathrm{decay,50\%-90\%}$, respectively. So these positive correlations are all very significant. We used a linear function in a logarithmic scale ($\log \tau = A+B\times\log E$)  to fit all of the six correlations with a Bayesian approach in \citet{kelly07}. Then we got $\log \tau_\mathrm{rise,10\%-90\%} =(-23.31\pm3.76)+(0.55\pm0.09)\times\log E$, $\log \tau_\mathrm{rise,10\%-50\%} =(-18.08\pm4.75)+(0.42\pm0.11)\times\log E$, $\log \tau_\mathrm{rise,50\%-90\%} =(-22.08\pm4.41)+( 0.52\pm0.10)\times\log E$, $\log \tau_\mathrm{decay,10\%-90\%} =(-26.49\pm2.49)+(0.63\pm0.06)\times\log E$, $\log \tau_\mathrm{decay,10\%-50\%} =(-26.19\pm3.02)+(0.62\pm0.07)\times\log E$, and $\log \tau_\mathrm{decay,50\%-90\%} =(-26.00\pm3.73)+(0.62\pm0.08)\times\log E$, respectively. The intrinsic scatters of all of the above six correlations  are 0.39$\pm$0.03, 0.47$\pm$0.04, 0.44$\pm$0.04, 0.22$\pm$0.02, 0.27$\pm$0.03, and 0.37$\pm$0.03 dex, respectively. The best-fit results are also plotted in \autoref{tau_te}. The correlation between $\tau_\mathrm{rise,10\%-90\%}$ and $E$ also holds in an individual source GX 339$-$4, and the correlation between $\tau_\mathrm{decay,10\%-90\%}$ and $E$ also holds in an individual source H1743$-$322 (see the upper panels of \autoref{tau_te}). We also found a marginal correlation between the total radiated energy $E$ and the orbital period $P_\mathrm{orb}$ among different sources (see \autoref{orb_te}). The Spearman correlation coefficient is $0.49$ at a significance of $3.03 \sigma$. This correlation seems to only hold at an orbital period of less than about 100 hr because of the large deviation from this correlation of GRO J1744$-$28 with orbital period $~$284 hr. We need more data to test whether the correlation is valid at orbital periods larger than 100 hr. 

For the sources with unknown distances, we estimated the total fluence for each available outburst. If these outbursts also follow the correlations between $E$ and $\tau_\mathrm{rise}$ or $\tau_\mathrm{decay}$ shown in \autoref{tau_te}, we can infer the distances for these sources according to the correlations. We used a linear model in a logarithmic scale ($\log E = A+B\times\log \tau$) to fit the correlations with a Bayesian approach in \citet{kelly07}. The best-fit parameters are shown in \autoref{ab}. There are five sources with unknown distances in our sample (see \autoref{parameters}). We identified three outbursts in SLX 1746$-$331 and one outburst in four other sources (SWIFT J1539.2$-$6227, SWIFT J1842.5$-$1124, XTE J1755$-$324, and XTE J2012+381). The $e$-folding rise or decay timescales of these five sources can be seen in \autoref{results}. We calculated six values for the distance at most for each outburst according to the fitted results of different $e$-folding rise or decay timescales in \autoref{ab} by assuming that those outbursts follow the correlations.  We used the average value and standard deviations of the distances from the different correlations as our estimation. The distances for these five sources are $10.81\pm3.52$, $9.32\pm1.53$, $12.87\pm2.32$, $11.09\pm4.52$, and $7.49\pm2.16$ kpc for SLX 1746$-$331, SWIFT J1539.2$-$6227, SWIFT J1842.5$-$1124, XTE J1755$-$324, and XTE J2012+381, respectively. 

\begin{figure}
\centering
\includegraphics[width=\linewidth]{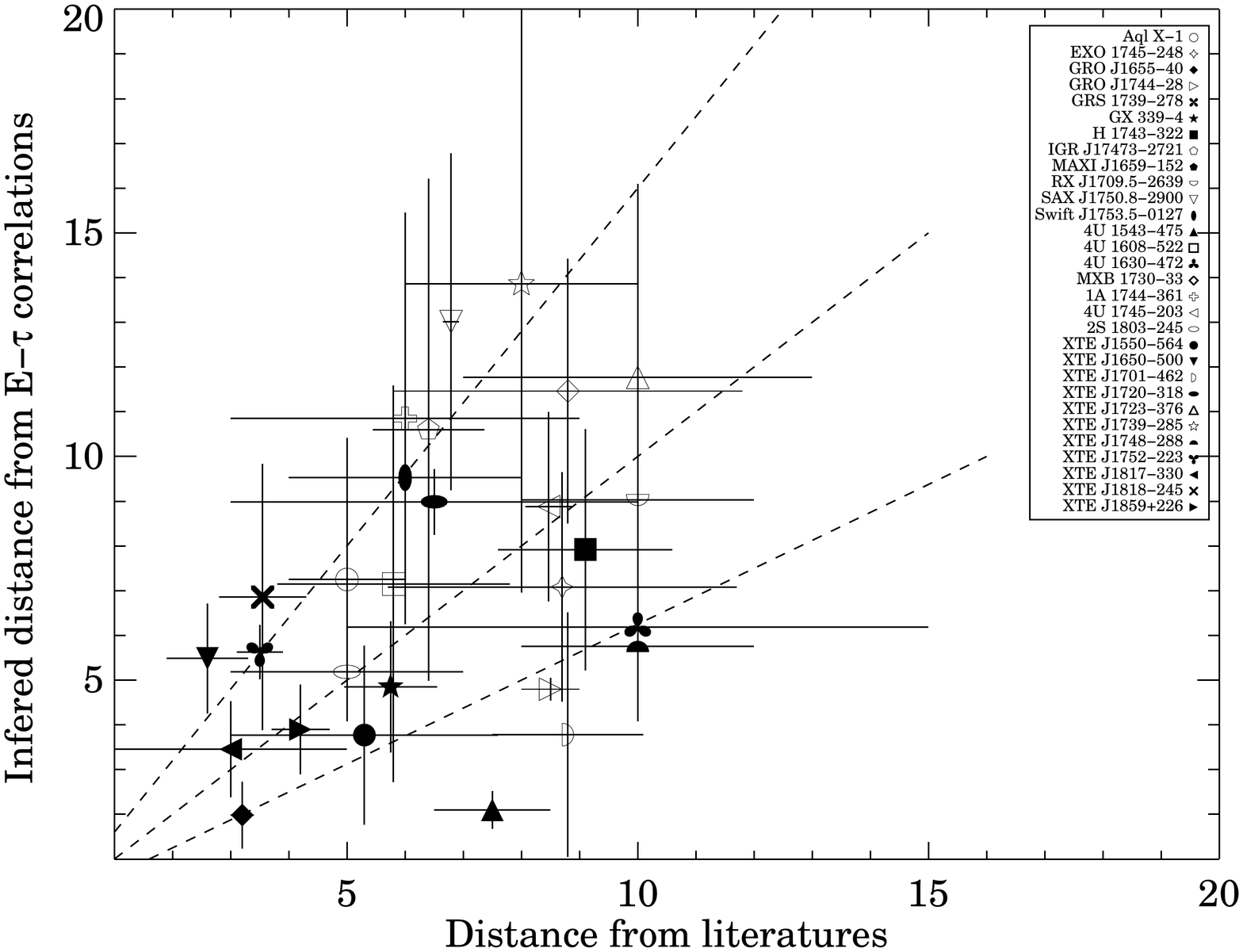}
\caption{Distances from literature versus distances inferred from $E-\tau$ correlations. From top to bottom, the three dashed lines represent $Y=1.6X$, $Y=X$, and $Y=\frac{X}{1.6}$, respectively. The differences between the inferred distances and the known distances of most of the sources are consistent with the intrinsic scatters of the $E-\tau$ correlations. }
\label{dis}
\end{figure}

\begin{figure}
\centering
\includegraphics[width=\linewidth]{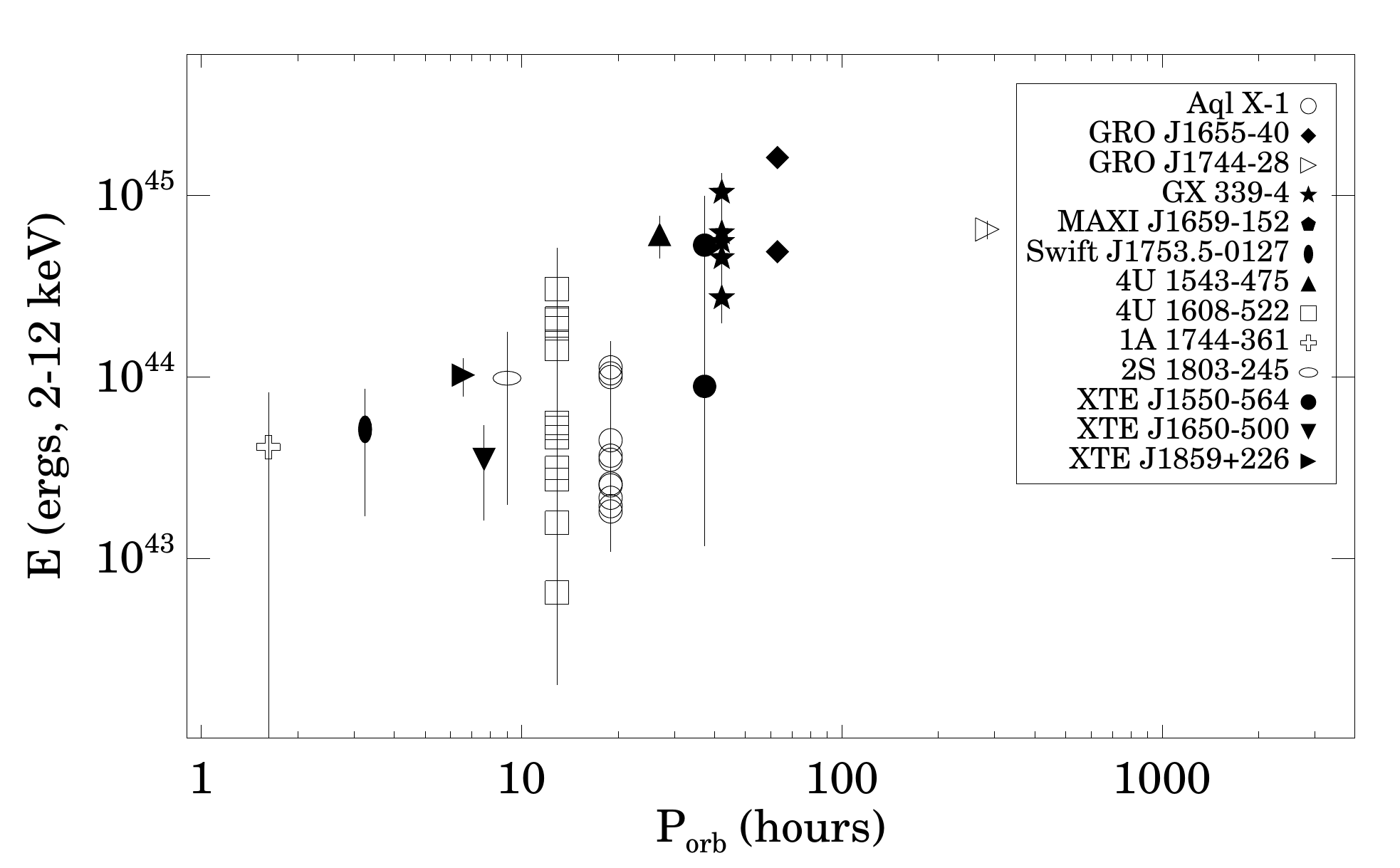}
\caption{Relation between orbital period and total radiated energy. The filled and unfilled
symbols represent BH and NS LMXBTs, respectively. There is a positive correlation between $E$ and $P_\mathrm{orb}$.}
\label{orb_te}
\end{figure}

The intrinsic scatters of the six different $E-\tau$ correlations are quite large, from about 0.3 to 0.5 dex. An average intrinsic scatter of 0.4 dex implies that the 1$\sigma$ uncertainties of the distances inferred according to those correlations can be up to a factor of 1.6. We further applied this method to sources with known distances and compared the inferred distances with the distances we collected from the literature (see \autoref{dis}). The differences between these two quantities of most of the sources are less than a factor of 1.6, which supports the idea that the large scatter of \autoref{dis} is due to the intrinsic scatters in the $E-\tau$ correlations. A few sources such as 4U 1543$-$47 show a large discrepancy, which is caused by the large deviation from the $E-\tau$ correlations of this source.

\capstartfalse
\begin{deluxetable}{lcc}
\centering
\tablecolumns{3}
\tabletypesize{\footnotesize}
\tablecaption{ Parameters of the Best-fit Results of $\log E = A+B\times\log \tau$}
\tablewidth{0pt}
\tablehead{
 \colhead{}  & \colhead{A} &\colhead{B} 
}
\startdata
$\tau_\mathrm{rise,10\%-90\%}$ vs. $E$ & $43.46 \pm 0.10$ & $0.68 \pm 0.11$ \\ 
$\tau_\mathrm{rise,10\%-50\%}$ vs. $E$ & $43.72 \pm 0.10$ & $0.49 \pm 0.13$ \\
$\tau_\mathrm{rise,50\%-90\%}$ vs. $E$ & $43.48 \pm 0.12$ & $0.60 \pm 0.12$ \\
$\tau_\mathrm{decay,10\%-90\%}$ vs. $E$ & $42.65 \pm 0.12$ & $1.14 \pm 0.10$ \\
$\tau_\mathrm{decay,10\%-50\%}$ vs. $E$ & $42.92 \pm 0.12$ & $1.00 \pm 0.11$ \\
$\tau_\mathrm{decay,50\%-90\%}$ vs. $E$ & $43.09 \pm 0.13$ & $0.75 \pm 0.10$ \\
\enddata
\label{ab}
\end{deluxetable}
\capstarttrue

\subsection{Summary of the Results}
We have performed a systematic study of the LMXBT outburst properties in the 2--12 keV band using the data from {\it RXTE}/ASM. We defined parameters that are used to describe outburst properties and presented the statistical results of these parameters, including peak X-ray luminosity, rate of change  of luminosity during the rise or decay phase on a daily timescale, $e$-folding rise or decay timescale, outburst duration, and total radiated energy. Readers need to keep in mind that these parameters depend on the energy band: all of our measurements are in the 2--12 keV band. The bolometric luminosity and total radiated energy could be three times larger than what we measured in 2--12 keV \citep[e.g. ][]{zand07}. Please note that our sample only includes the bright outbursts with a peak count rate larger than 0.1 Crab.
\begin{enumerate}
\item The average peak X-ray luminosity of outbursts in our sample is about 4.7$\times$10$^{37}$ ergs s$^{-1}$ (see $L_\mathrm{peak}$ in \autoref{lpeak}).
\item The average rate of change  of luminosity is about 0.01 $L_\mathrm{Edd}$/day during the rise phase (see $\dot L_\mathrm{rise,10\%-90\%}$ in \autoref{ldot}) and 0.004 $L_\mathrm{Edd}$/day during the decay phase (see $\dot L_\mathrm{decay,10\%-90\%}$ in \autoref{ldot}). 
\item The average $e$-folding rise timescale is about five days (see $\tau_\mathrm{rise,10\%-90\%}$ in \autoref{tau}), and the $e$-folding decay timescale is about 15 days (see $\tau_\mathrm{decay,10\%-90\%}$ in \autoref{tau}). 
\item  The average outburst duration is about 54 days (see \autoref{t_dur}).
\item The average total radiated energy of all LMXBTs is about 0.91$\times$10$^{44}$ ergs  for the bright outbursts (see $<E_\mathrm{total}>$ in \autoref{t_te}).
\end{enumerate}

We have also found correlations between some outburst properties or between outburst properties and binary system parameters:
\begin{enumerate}
\item A positive correlation between the rate of change  of luminosity and the peak X-ray luminosity in units of $L_\mathrm{Edd}$ in both the rise and decay phases (see \autoref{ldot_peak}).
\item  A positive correlation between the peak X-ray luminosity and the total radiated energy (see \autoref{p_te}).
\item  A positive correlation between the $e$-folding rise or decay timescale and the total radiated energy (see \autoref{tau_te}).
\item  A weak positive correlation between the orbital period and the total radiated energy (see \autoref{orb_te}).
\item A weak positive correlation between the orbital period and the $e$-folding rise timescale (see \autoref{orb_rise}). 
\end{enumerate}

\section{DISCUSSION}
\subsection{Comparison between Outbursts of BH and NS LMXBTs}
Once a new LMXBT was discovered, the most important thing is to determine the nature of the compact object. If there is no evidence of an NS from a Type I burst or pulsation detection or if the mass of the compact object cannot be measured, then people normally use the X-ray spectral and timing properties to infer the nature of the compact object.

Our statistical study of the outbursts of LMXBTs shows that the average values of most parameters have significant differences between BH LMXBTs and NS LMXBTs, except $\tau_\mathrm{rise,10\%-50\%}$. 
\begin{enumerate}
\item The average peak count rate and luminosity of BH LMXBTs is about two times larger than that of NS LMXBTs.
\item The average $e$-folding rise timescales (except $\tau_\mathrm{rise,10\%-50\%}$) of BH LMXBTs are about two times larger than those of NS LMXBTs.
\item The average $e$-folding decay timescale of BH LMXBTs are all more than two times larger than those of NS LMXBTs.
\item The average outburst duration of BH LMXBTs is about two times larger than that of NS LMXBTs.
\item The average total radiated energy of BH LMXBTs is about five times larger than that of NS LMXBTs.
\end{enumerate}

Therefore, these parameters are very helpful in inferring the nature of the compact object in a statistical sense. When a new transient is discovered, we can measure the outburst properties (including peak X-ray luminosity, rise or decay timescale, outburst duration, and total radiated energy) as described in Section 2, and compare them with the distributions of all of the available parameters of BH LMXBTs and NS LMXBTs to jointly infer the nature of the central compact object. 

\subsection{Disk Mass as the Primary Initial Condition for Nonstationary Accretion}
\citet{yy09} have suggested that the nonstationary accretion that is characterized by the rate of increase of X-ray luminosity plays an important role in determining the luminosity of the hard-to-soft state transition and the peak X-ray luminosity of an outburst. In order to further investigate the role of nonstationary accretion, we studied the rate of change of luminosity in the rise and decay phases. The rate of change  of luminosity in the rise phase is about 2 times larger than that in the decay phase (both for BH and NS LMXBTs). The correlation between the rate of change  of luminosity and peak X-ray luminosity scaled in $L_\mathrm{Edd}$ exists both in the rise and decay phase (see \autoref{ldot_peak}), which confirms the previous results and further supports the schematic picture of Fig.28 in \citet{yy09}.

The nonstationary accretion corresponding to transient outbursts is probably set up by an initial condition: disk mass \citep{yu04,yu07,wu10a,wu10}. In our study, the total radiated energy $E$ should correspond to the disk mass accreted during an outburst if a constant radiative efficiency can be approximately assumed. There are three outbursts followed by immediate secondary outbursts in our sample, including the 2002 and 2010 outbursts of 4U 1630$-$47 and the 1998 outburst of XTE J1550$-$564. Based on our definition, the $E$ only includes the radiated energy during the primary outburst in these cases. But only three data points cannot erase the positive correlations related with E. Thus we concluded that the disk mass accreted during an outburst is correlated with peak X-ray luminosity, rise timescale, and decay timescale, which supports the idea that disk mass plays a major role in determining outburst properties inferred from observations \citep{yu04,yu07,wu10a,wu10}. The peak X-ray luminosity is a function of disk mass, which is actually expected from the DIM \citep{king98,shahbaz98,lasota01,dubus01}, because the peak mass accretion rate $\dot{M}_\mathrm{peak}$ depends on the maximum radius over which the heating front can propagate. 

We found a weak correlation between the total radiated energy $E$ and the orbital period $P_\mathrm{orb}$ (see \autoref{orb_te}). For a fixed mass ratio between the compact star and the companion star, the orbital period constrains the size of the Roche lobe and the maximum disk size as well. Only if a specific ratio of the Roche lobe was filled by the accretion disk, and a specific ratio of mass stored in the disk was accreted onto the compact star during an outburst, can we see such a positive correlation between $E$ and $P_\mathrm{orb}$. This correlation is also expected from the DIM \citep{lasota01} because the mass accreted during an inside-out outburst is mainly determined by the disk radius where the heating front can reach; the critical surface density change very small between different systems. If the irradiation is strong enough to ionize the whole disk, the heat front can propagate to the outer radius, and most of the disk mass will be accreted by the compact object during the outburst \citep{king98,shahbaz98,dubus01}. The average peak X-ray luminosity of our sample is about 4.7 $\times$ 10$^{37}$ ergs s$^{-1}$ (see \autoref{lpeak}), which is larger than the critical luminosity required to ionize the entire disk for a typical disk size given in \citet{shahbaz98}, so we can see that the correlation between the total radiated energy $E$ and the orbital period $P_\mathrm{orb}$. GRO J1744$-$28 has the longest orbital period in our sample, which means the largest disk size. A possible reason for GRO J1744$-$28 showing a large deviation from the correlation (see \autoref{orb_te}) is that the peak X-ray luminosity is not high enough to ionize the whole disk. 

\subsection{Transient Stellar-mass ULXs During Super-Eddintong Outbursts}
A statistical study of the hard-to-soft state transitions in Galactic bright X-ray binaries shows no indication of a luminosity saturation or cutoff in the correlation between the hard-to-soft transition luminosity and the peak X-ray luminosity of the outburst or flare up to more than 30\% of $L_\mathrm{Edd}$ \citep{yy09}. This indicates that more luminous hard states and outbursts than those observed in our Galaxy are allowed by physics. If the luminosity function of Galactic LMXBT outbursts can be estimated, we would see that one outburst of Galactic LMXBTs can reach an Eddington outburst in a few tens of years \citep{yy09}. A super-Eddington outburst candidate is the 1999 outburst of V4641 Sgr \citep[see][]{Orosz2001,Revnivtsev2002}. However, this source was classified as HMXB in the subsequent studies \citep{Orosz2001,Liu2006}. On the other hand, the 1999 outburst of V4641 Sgr shows properties that are largely different from the classic outburst of an LMXBT. The rise and decay phases of this outburst were extremely fast: it reached 12 Crab within 8 hr from quiescence and returned to quiescence within 2 hr. We speculated that this outburst has a mechanism different from the classic outburst of an LMXBT.

Two transient ULXs in M31 were identified as LMXBs \citep{henze12,kaur12,midd12} which confirmed our previous prediction of the occurrence of transient ULXs in \citet{yy09}, because M31 is significantly larger than the Milky Way. Monitoring observations performed with $Chandra$, $XMM-Newton$ and $Swift$ covered the entire outburst phase for XMMU J004243.61+412519 and the decay phase of CXOM31 J004253.1+411422. We collected the X-ray light curves of both transient ULXs from the literature \citep{kaur12,mid13} in order to compare the outburst properties with the Galactic LMXBTs. We followed the same methods in Section 2 to measure the peak X-ray luminosity, rate of change  of luminosity, $e$-folding rise and decay timescales, outburst duration, and total radiated energy (see \autoref{m31}). The light curves of these two ULXs were obtained in 0.3--10 keV, which is different from the {\it RXTE}/ASM light curves of Galactic LMXBTs. Comparing the parameters of the outbursts of the two transient ULXs in M31 (see \autoref{m31}) with NS LMXBTs and BH LMXBTs, we tend to suggest that the nature of the central compact objects is like BHs. The masses of central compact objects in these two ULXs are both assumed to be 10 $M_{\sun}$.

The outburst parameters of the two M31 transient ULXs are also plotted in the correlations we found in the Galactic LMXBTs. As can be seen in \autoref{ldot_peak} and \autoref{p_te}, the two transient ULXs in M31 follow the correlation between the rate of change  of luminosity and the peak X-ray luminosity and the correlation between the peak X-ray luminosity and the total radiated energy. But as shown in see \autoref{tau_te}, the two transient ULXs seem to locate below the extrapolation of the correlations between the $e$-folding rise or decay timescale and the total radiated energy found in Galactic LMXBTs. Because of the large intrinsic scatters of the correlations, we need more data to verify whether the correlations have a saturation or break at the higher luminosity end. In the future, much more sensitive all-sky monitors will provide us with a large sample of light curves of transient ULXs in nearby galaxies, which will offer a significant advantage ing studying the properties of super-Eddington outbursts because these sources have more populations and smaller absorption compared to Galactic sources and shorter timescales compared to super-Eddington active galactic nuclei. These two transient ULXs in M31 roughly follow the correlations found in the Galactic LMXBTs, and their outburst properties are also similar to the Galactic LMXBTs, which demonstrates that the sub-Eddington and super-Eddington outbursts are driven by the same mechanism.

\subsection{Rise or Decay Timescale and the DIM} 
 In the original form of the DIM, the rise or decay time corresponds to the time duration over which the heating or cooling front propagates. The velocity of the heating front is of the order of $\alpha c_\mathrm{s}$ \citep{meyer84}, where $c_\mathrm{s}$ is the sound speed, so the rise time $t_\mathrm{rise}$ is $\sim R/\alpha c_\mathrm{s}$ \citep{frank02}. If the rise time corresponds to the propagation time through the entire accretion disk, it will be $\sim$ 100 hr if assuming a disk size of $\sim 10^{11}$ cm, which is comparable with the rise timescale we observed ($\tau_\mathrm{rise,10\%-90\%} \sim 5$ days; see the Table.~\ref{tau}). But the rise timescale we measured is in the X-ray band, so it may correspond to the heating front propagation time through the X-ray emission region, which is much smaller.

\citet{dubus01} considered a modified DIM, including the effects of irradiation and evaporation, and assumed that the inner region of a disk is replaced by an advection-dominated accretion flow (ADAF) at the beginning of an outburst. The inner radius of the truncated disk decreases until its minimum value is reached when $\dot{M}_\mathrm{in}=\dot{M}_\mathrm{ev}(R_\mathrm{min})$, where $\dot{M}_\mathrm{in}$ is the mass accretion rate and $\dot{M}_\mathrm{ev}$ is  the evaporation rate. The rise time $t_\mathrm{rise}$ is a viscous time that the disk needs to reach $R_\mathrm{min}$ from the initial truncated radius $R_\mathrm{tr}$. For the typical parameters of LMXBs, the $t_\mathrm{rise}$ is of the order of several days \citep{dubus01}, which is comparable with the rise timescale we observed. But we did not measure the rise timescale by taking the quiescence as the beginning. In practice, the rise time from quiescence is difficult to estimate from the {\it RXTE}/ASM data because the sensitivity is not enough to determine the quiescent flux level.

Our statistical results show that the outburst decay of most outbursts is consistent with an exponential form down to $F_{10\%}$. It has been found that the irradiation of the accretion disk by the central X-rays can naturally explain the exponential decay \citep{king98}. The decay timescale ($\sim$20--40 days) given by \citet{king98} is comparable to the average decay timescale we measured within error bars (6 -- 37 days; see $\tau_\mathrm{decay,10\%-90\%}$ in \autoref{tau}). \citet{king98} also showed that the $e$-folding decay timescale is of the order of the viscous timescale of the outer radius and increases with the orbital period, but saturates at $\tau_\mathrm{decay} \sim$ 40 days for the binary systems with orbital periods longer than about a day. We have checked for this but did not find a positive correlation between orbital period and $e$-folding decay timescale for the data where $\tau_\mathrm{decay}$ is below 40 days. The broadband spectral evolution (including optical, ultraviolet, and X-ray bands) shows that the outer disk radius decreases during the outburst decay \citep[see, e.g.,][]{Hynes2002}. So the $\tau_\mathrm{decay}$ as a proxy for the viscous timescale of the outer disk radius cannot trace the size of the Roche lobe. This might be the reason that we did not find a statistical positive correlation between  $\tau_\mathrm{decay}$ and the orbital period.

However, we found a positive correlation between the $e$-folding rise timescale and the orbital period (see \autoref{orb_rise}). It is reasonable that the viscous time at the outer disk is positively correlated with the orbital period \citep{menou99,gilfanov05}, so the rise timescale may relate to the viscous time of the outer disk. If the $e$-folding rise timescale we measured is a proxy for the viscous timescale at the initial truncated radius, the truncated radius must be related to the outer disk radius, which leads to the correlation between the $e$-folding rise timescale and the orbital period. How the truncated radius is related to the outer disk radius when the LMXBTs are in their quiescence deserves further studies.

\section{Summary and Conclusion}
We have performed a statistical study of the outburst properties of 110 outbursts in 36 LMXBTs.  We have measured the average and standard deviation of the outburst parameters, including the peak X-ray luminosity, rate of change  of luminosity during the rise or decay phase on a daily timescale, e-folding rise or decay timescale, outburst duration, and total radiated energy. We have found the following:
\begin{enumerate}
\item  Most of the outburst parameters show significant differences between BH and NS LMXBTs, which is helpful in determining the nature of the compact object in a newly discovered LMXBTs. 
\item   A positive correlation exists between the outburst peak X-ray luminosity and the rate of change  of X-ray luminosity in both the rise and the decay phases, which is consistent with our previous studies \citep{yy09}. 
\item  Positive correlations between the total radiated energy and peak X-ray luminosity, and between the total radiated energy and the $e$-folding rise or decay timescale in the outbursts, which implies that the disk mass before an outburst is the primary initial condition that sets up the outburst properties.
\item  The two transient stellar-mass ULXs in M31 roughly follow the correlations we found in Galactic LMXBTs, which indicates that they share a common outburst mechanism with Galactic LMXBTs. 
\end{enumerate}

\acknowledgments
We are grateful to the anonymous referee. We would like to thank Jean-Pierre Lasota and Diego Altamiratno for the very helpful comments. This work was supported in part by the National Natural Science Foundation of China under grant No. 11403074, 11333005, and 11350110498, by Strategic Priority Research Program "The Emergence of Cosmological Structures" under Grant No. XDB09000000, and the XTP project under Grant No. XDA04060604, by the Shanghai Astronomical Observatory Key Project, and by the Chinese Academy of Sciences Fellowship for Young International Scientists Grant. Zhen Yan acknowledges the support from the Knowledge Innovation Program of the Chinese Academy of Sciences. The study has made use of data obtained through the High Energy Astrophysics Science Archive Research Center Online Service, provided by the NASA Goddard Space Flight Center.

\input{yz.bbl}

\input{ltab0.tex}

\input {ltab1.tex}

\end{document}

%% file: ltab0.tex
\clearpage
\LongTables
\begin{landscape}

\tabletypesize{\tiny}
\begin{deluxetable}{lcccccccccc}

\tablecaption{The Parameters of LMXBT Outbursts \label{results}}
\tablewidth{0pt}
\tablehead{
\colhead{Source} &\colhead{$t_\mathrm{peak}$}
&\colhead{$F_\mathrm{peak}$ }
&\colhead{$\tau_\mathrm{rise,10\%-90\%}$} &\colhead{$\tau_\mathrm{rise,10\%-50\%}$} 
&\colhead{$\tau_\mathrm{rise,50\%-90\%}$} &\colhead{$\tau_\mathrm{decay,10\%-90\%}$}
&\colhead{$\tau_\mathrm{decay,10\%-50\%}$} &\colhead{$\tau_\mathrm{decay,50\%-90\%}$}
&\colhead{Outburst Duration} &\colhead{$E$} \\
\colhead{} &\colhead{MJD}
&\colhead{(crab)}
&\colhead{(days)} &\colhead{(days)} 
&\colhead{(days)} &\colhead{(days)}
&\colhead{(days)} &\colhead{(days)}
&\colhead{(days)} &\colhead{($10^{44}$ ergs)}
}
\startdata
Swift J1539.2$-$6227 & 54815 & 0.28$\pm$ 0.02 & 7.12$\pm$ 2.15 & 1.77$\pm$ 0.85 & 8.28$\pm$ 1.16 & 12.04$\pm$ 5.11 & 13.88$\pm$ 7.70 & 6.11$\pm$ 1.90 & 47.00 & \nodata \\
4U 1543$-$475 & 52445 & 3.79$\pm$ 0.02 & 0.79$\pm$ 0.01 & 0.34$\pm$ 0.01 & 2.37$\pm$ 0.02 & 8.82$\pm$ 0.11 & 8.73$\pm$ 0.15 & 9.08$\pm$ 0.23 & 27.00 & 6.11$\pm$   1.63 \\
XTE J1550$-$564 & 51075 & 6.52$\pm$ 0.04 & 4.36$\pm$ 0.02 & 6.12$\pm$ 0.04 & 1.21$\pm$ 0.01 & 19.17$\pm$ 0.13 & 28.19$\pm$ 0.27 & 2.65$\pm$ 0.03 & 51.00 & 5.32$\pm$   4.62  \\
 & 51663 & 0.95$\pm$ 0.01 & 9.07$\pm$ 0.17 & 11.95$\pm$ 0.40 & 3.09$\pm$ 0.15 & 8.66$\pm$ 0.48 & 5.68$\pm$ 0.51 & 18.20$\pm$ 3.20 & 41.00 & 0.89$\pm$   0.77 \\
4U 1630$-$472 & 50247 & 0.34$\pm$ 0.02 & 46.20$\pm$ 4.40 & 3.21$\pm$ 0.50 & 115.97$\pm$ 12.51 & 22.59$\pm$ 1.38 & 14.72$\pm$ 1.03 & 44.64$\pm$ 6.64 & 145.00 & 7.00$\pm$   7.00 \\
 & 50864 & 0.37$\pm$ 0.01 & 5.38$\pm$ 0.22 & 2.33$\pm$ 0.12 & 15.49$\pm$ 0.75 & 16.15$\pm$ 0.67 & 13.97$\pm$ 1.07 & 20.78$\pm$ 2.21 & 51.00 & 2.13$\pm$   2.13  \\
 & 51359 & 0.25$\pm$ 0.02 & 24.02$\pm$ 4.05 & 13.20$\pm$ 2.78 & 64.58$\pm$ 12.23 & 11.30$\pm$ 0.98 & 7.68$\pm$ 1.67 & 18.83$\pm$ 7.60 & 78.00 & 2.25$\pm$   2.25 \\
 & 51884 & 0.56$\pm$ 0.05 & 5.09$\pm$ 0.60 & 1.20$\pm$ 0.17 & 27.20$\pm$ 6.49 & 18.68$\pm$ 1.13 & 17.08$\pm$ 1.84 & 21.26$\pm$ 3.17 & 72.00 & 4.62$\pm$   4.62  \\
 & 52627 & 0.84$\pm$ 0.04 & 38.57$\pm$ 1.33 & 2.65$\pm$ 0.11 & 8.26$\pm$ 0.77 & 19.43$\pm$ 2.27 & 17.40$\pm$ 2.80 & 24.53$\pm$ 3.77 & 140.00 & 11.39$\pm$  11.39 \\
 & 53747 & 0.31$\pm$ 0.02 & 19.36$\pm$ 6.50 & 9.62$\pm$ 4.06 & 55.40$\pm$ 18.98 & 44.94$\pm$ 5.99 & 42.18$\pm$ 8.66 & 52.50$\pm$ 14.70 & 146.00 & 5.52$\pm$   5.52  \\
 & 54479 & 0.26$\pm$ 0.02 & 7.53$\pm$ 5.45 & 5.21$\pm$ 4.59 & 18.43$\pm$ 6.31 & 61.45$\pm$ 9.13 & 29.75$\pm$ 5.97 & 151.02$\pm$ 15.08 & 167.00 & 6.55$\pm$   6.55 \\
 & 55224 & 0.32$\pm$ 0.02 & 10.13$\pm$ 2.35 & 2.66$\pm$ 0.88 & 29.91$\pm$ 7.42 & 11.27$\pm$ 1.24 & 5.54$\pm$ 0.98 & 23.31$\pm$ 4.25 & 53.00 & 2.76$\pm$   2.76 \\
XTE J1650$-$500 & 52172 & 0.56$\pm$ 0.00 & 7.21$\pm$ 1.11 & 1.37$\pm$ 0.30 & 20.99$\pm$ 0.91 & 29.12$\pm$ 3.39 & 21.22$\pm$ 3.42 & 50.56$\pm$ 3.60 & 81.00 & 0.35$\pm$   0.19   \\
XTE J1652$-$453 & 55016 & 0.18$\pm$ 0.01 & 3.28$\pm$ 1.73 & 1.90$\pm$ 1.35 & 7.21$\pm$ 2.68 & 23.72$\pm$ 8.37 & 24.22$\pm$ 12.04 & 22.45$\pm$ 6.67 & 61.00 & \nodata \\
GRO J1655$-$40 & 50284 & 3.21$\pm$ 0.02 & 37.00$\pm$ 0.95 & 14.54$\pm$ 0.53 & 91.69$\pm$ 1.24 & 172.34$\pm$ 1.47 & 49.93$\pm$ 0.66 & 168.22$\pm$ 4.05 & 461.00 & 16.18$\pm$   2.02 \\
 & 53508 & 4.31$\pm$ 0.03 & 22.87$\pm$ 0.32 & 4.11$\pm$ 0.08 & 5.72$\pm$ 0.08 & 51.84$\pm$ 0.41 & 62.40$\pm$ 0.65 & 20.41$\pm$ 0.43 & 171.00 & 4.90$\pm$   0.61\\
MAXI J1659$-$152 & 55476 & 0.63$\pm$ 0.04 & \nodata & \nodata & 1.40$\pm$ 0.17 & 7.24$\pm$ 2.21 & 9.57$\pm$ 4.13 & 1.48$\pm$ 0.23 & \nodata & \nodata \\
GX 339$-$4 & 50867 & 0.33$\pm$ 0.02 & 46.65$\pm$ 5.34 & 34.42$\pm$ 5.99 & 78.01$\pm$ 14.54 & 114.78$\pm$ 34.43 & 27.48$\pm$ 11.09 & 372.52$\pm$ 61.78 & 432.00 & 6.22$\pm$   1.73 \\
 & 52482 & 0.88$\pm$ 0.01 & 49.03$\pm$ 1.59 & 20.66$\pm$ 0.92 & 63.21$\pm$ 2.39 & 78.13$\pm$ 3.03 & 77.87$\pm$ 4.18 & 78.83$\pm$ 3.40 & 317.00 & 10.41$\pm$   2.90 \\
 & 53310 & 0.53$\pm$ 0.02 & 39.25$\pm$ 5.50 & 27.72$\pm$ 5.25 & 72.10$\pm$ 2.53 & 64.08$\pm$ 6.79 & 60.87$\pm$ 9.31 & 73.27$\pm$ 12.25 & 273.00 & 4.54$\pm$   1.26 \\
 & 54149 & 1.03$\pm$ 0.02 & 14.19$\pm$ 0.82 & 16.45$\pm$ 1.29 & 8.15$\pm$ 0.42 & 25.04$\pm$ 0.79 & 18.24$\pm$ 0.90 & 42.91$\pm$ 2.77 & 95.00 & 2.73$\pm$   0.76 \\
 & 55317 & 0.66$\pm$ 0.02 & 21.33$\pm$ 1.59 & 18.11$\pm$ 1.94 & 30.77$\pm$ 4.08 & 139.03$\pm$ 36.04 & 139.03$\pm$ 46.65 & 139.03$\pm$ 15.91 & 382.00 & 5.57$\pm$   1.55 \\
XTE J1720$-$318 & 52652 & 0.44$\pm$ 0.01 & 21.30$\pm$ 13.28 & \nodata & \nodata & 24.88$\pm$ 3.98 & 23.17$\pm$ 5.77 & 30.10$\pm$ 12.81 & 113.00 & 0.98$\pm$   1.06 \\
GRS 1739$-$278 & 50159 & 0.84$\pm$ 0.02 & \nodata & \nodata & 3.64$\pm$ 0.20 & 84.36$\pm$ 4.59 & 97.32$\pm$ 7.25 & 48.70$\pm$ 3.95 & \nodata & \nodata \\
H 1743$-$322 & 52752 & 1.39$\pm$ 0.01 & 9.11$\pm$ 0.19 & 10.32$\pm$ 0.33 & 6.20$\pm$ 0.20 & 76.14$\pm$ 18.18 & 72.85$\pm$ 23.87 & 21.74$\pm$ 1.77 & 179.00 & 15.67$\pm$   5.17 \\
 & 53234 & 0.33$\pm$ 0.02 & 19.97$\pm$ 4.52 & 7.54$\pm$ 2.31 & 55.15$\pm$ 10.46 & 17.23$\pm$ 1.19 & 13.36$\pm$ 1.39 & 26.99$\pm$ 2.93 & 96.00 & 3.12$\pm$   1.03 \\
 & 53602 & 0.20$\pm$ 0.01 & 3.75$\pm$ 0.86 & 2.45$\pm$ 0.73 & 8.05$\pm$ 1.94 & 11.64$\pm$ 7.99 & 10.49$\pm$ 9.73 & 14.93$\pm$ 1.69 & 42.00 & 0.70$\pm$   0.23 \\
 & 54466 & 0.33$\pm$ 0.04 & 9.43$\pm$ 3.71 & 14.05$\pm$ 8.92 & 1.36$\pm$ 0.44 & 13.51$\pm$ 2.24 & 13.87$\pm$ 2.94 & 12.21$\pm$ 4.02 & 53.00 & 1.36$\pm$   0.45 \\
 & 54988 & 0.25$\pm$ 0.01 & 5.24$\pm$ 1.28 & 6.22$\pm$ 2.20 & 2.93$\pm$ 0.71 & 12.98$\pm$ 6.12 & 14.73$\pm$ 9.99 & 8.65$\pm$ 3.29 & 41.00 & 0.72$\pm$   0.24 \\
 & 55425 & 0.22$\pm$ 0.01 & 3.20$\pm$ 13.69 & 2.86$\pm$ 14.18 & 5.35$\pm$ 1.84 & 7.94$\pm$ 7.07 & 8.83$\pm$ 11.10 & 5.77$\pm$ 2.80 & 37.00 & 0.64$\pm$   0.21  \\
XTE J1748$-$288 & 50970 & 0.57$\pm$ 0.01 & 1.35$\pm$ 0.25 & 1.18$\pm$ 0.25 & 2.99$\pm$ 0.24 & 14.45$\pm$ 1.63 & 9.87$\pm$ 1.43 & 30.73$\pm$ 1.70 & 37.00 & 2.32$\pm$   0.93 \\
SLX 1746-331  &       52740  &   0.37$\pm$   0.01  &   2.78$\pm$   0.21  &   2.28$\pm$   0.28  &   3.55$\pm$   0.20  &  25.09$\pm$   4.09  &  15.02$\pm$   3.51  &  49.32$\pm$   5.09  &  60.00 &    \nodata  \\
       &       54383  &   0.14$\pm$   0.01  &   1.72$\pm$   1.10  &   1.64$\pm$   1.16  &   2.43$\pm$   0.76  &  10.13$\pm$   3.04  &   8.26$\pm$   3.42  &  15.36$\pm$   4.55  &  30.00 &    \nodata\\
       &       54475  &   0.16$\pm$   0.02  &  13.36$\pm$   7.14  &   8.25$\pm$   6.00  &  29.94$\pm$  20.38  &  33.91$\pm$   6.49  &  28.17$\pm$   8.73  &  45.12$\pm$  12.87  & 107.00 &    \nodata  \\
XTE J1752$-$223 & 55219 & 0.52$\pm$ 0.02 & 37.67$\pm$ 6.50 & 29.99$\pm$ 6.32 & 58.74$\pm$ 22.33 & 33.66$\pm$ 4.49 & 32.01$\pm$ 5.81 & 38.69$\pm$ 8.78 & 161.00 & 1.07$\pm$   0.25 \\
Swift J1753.5$-$0127 & 53558 & 0.21$\pm$ 0.01 & 3.79$\pm$ 0.44 & 0.93$\pm$ 0.17 & 9.74$\pm$ 1.41 & 28.47$\pm$ 2.95 & 28.84$\pm$ 4.25 & 27.48$\pm$ 3.34 & 69.00 & 0.51$\pm$   0.34 \\
XTE J1755$-$324 & 50656 & 0.18$\pm$ 0.00 & 3.78$\pm$ 1.02 & 3.89$\pm$ 1.39 & 3.44$\pm$ 0.41 & 38.65$\pm$ 5.52 & 46.78$\pm$ 9.13 & 14.95$\pm$ 2.19 & 102.00 & \nodata \\
XTE J1817$-$330 & 53767 & 1.94$\pm$ 0.02 & 2.45$\pm$ 0.06 & 0.82$\pm$ 0.03 & 7.20$\pm$ 0.26 & 24.55$\pm$ 0.41 & 27.59$\pm$ 0.81 & 15.68$\pm$ 0.95 & 57.00 & 0.81$\pm$   1.08 \\
XTE J1818$-$245 & 53596 & 0.53$\pm$ 0.01 & 0.89$\pm$ 0.05 & 0.55$\pm$ 0.04 & 2.22$\pm$ 0.20 & 26.24$\pm$ 2.14 & 28.54$\pm$ 3.21 & 20.16$\pm$ 1.64 & 55.00 & 0.25$\pm$   0.11 \\
Swift J1842.5$-$1124 & 54737 & 0.15$\pm$ 0.01 & 7.79$\pm$ 2.18 & 8.62$\pm$ 3.83 & 6.23$\pm$ 1.52 & 30.43$\pm$ 14.59 & 28.57$\pm$ 19.32 & 34.98$\pm$ 5.54 & 89.00 & \nodata \\
XTE J1859+226 & 51467 & 1.31$\pm$ 0.01 & 3.30$\pm$ 0.03 & 3.42$\pm$ 0.05 & 3.02$\pm$ 0.08 & 17.16$\pm$ 0.38 & 21.42$\pm$ 0.70 & 5.84$\pm$ 0.20 & 50.00 & 1.02$\pm$   0.24 \\
XTE J2012+381 & 50967 & 0.21$\pm$ 0.00 & 2.65$\pm$ 0.19 & 1.98$\pm$ 0.26 & 3.44$\pm$ 0.28 & 31.90$\pm$ 5.26 & 18.19$\pm$ 4.04 & 19.43$\pm$ 1.62 & 84.00 & \nodata \\
\hline

4U 1608$-$522 & 50094 & 0.60$\pm$ 0.01 & \nodata & \nodata & 4.14$\pm$ 0.37 & 25.51$\pm$ 1.08 & 4.63$\pm$ 0.29 & 69.08$\pm$ 2.87 & \nodata & \nodata \\
 & 50850 & 0.95$\pm$ 0.01 & 1.44$\pm$ 0.07 & 1.18$\pm$ 0.07 & 2.64$\pm$ 0.14 & 21.26$\pm$ 0.45 & 22.23$\pm$ 0.67 & 18.74$\pm$ 0.38 & 54.00 & 1.86$\pm$   1.28  \\
 & 51896 & 0.42$\pm$ 0.03 & \nodata & \nodata & \nodata & 11.32$\pm$ 1.39 & 12.82$\pm$ 2.84 & 8.75$\pm$ 1.71 & \nodata & \nodata \\
 & 52220 & 0.65$\pm$ 0.02 & 2.10$\pm$ 0.52 & 0.99$\pm$ 0.35 & 4.75$\pm$ 0.46 & 4.21$\pm$ 0.50 & \nodata & \nodata & 19.00 & 0.56$\pm$   0.39\\
 & 52482 & 0.79$\pm$ 0.01 & 3.51$\pm$ 0.17 & 2.24$\pm$ 0.14 & 8.11$\pm$ 0.27 & 21.62$\pm$ 0.46 & 11.55$\pm$ 0.50 & 51.63$\pm$ 5.17 & 64.00 & 2.07$\pm$   1.42  \\
 & 52711 & 0.34$\pm$ 0.01 & 1.49$\pm$ 0.09 & 1.19$\pm$ 0.14 & 1.99$\pm$ 0.24 & 6.29$\pm$ 0.57 & 7.19$\pm$ 0.74 & 2.52$\pm$ 0.55 & 22.00 & 0.27$\pm$   0.19 \\
 & 53084 & 0.26$\pm$ 0.01 & 2.63$\pm$ 1.31 & 1.49$\pm$ 0.98 & 6.18$\pm$ 1.31 & 2.74$\pm$ 0.22 & 1.36$\pm$ 0.15 & 5.61$\pm$ 0.63 & 14.00 & 0.16$\pm$   0.11 \\
 & 53192 & 0.12$\pm$ 0.06 & 6.57$\pm$ 2.12 & 8.70$\pm$ 4.50 & 2.95$\pm$ 0.36 & 3.02$\pm$ 0.71 & 3.58$\pm$ 1.42 & 2.18$\pm$ 0.28 & 19.00 & 0.06$\pm$   0.04  \\
 & 53494 & 0.67$\pm$ 0.02 & 26.61$\pm$ 5.76 & 2.56$\pm$ 0.82 & 8.93$\pm$ 1.86 & 19.75$\pm$ 0.99 & 18.95$\pm$ 1.42 & 21.51$\pm$ 0.89 & 100.00 & 3.05$\pm$   2.11 \\
 & 53654 & 0.27$\pm$ 0.00 & 7.66$\pm$ 1.35 & 9.30$\pm$ 2.55 & 4.61$\pm$ 0.33 & 12.24$\pm$ 3.43 & 11.36$\pm$ 4.45 & 14.49$\pm$ 1.77 & 58.00 & 0.52$\pm$   0.36 \\
 & 54273 & 0.57$\pm$ 0.01 & 3.53$\pm$ 0.26 & \nodata & \nodata & 5.50$\pm$ 0.33 & 3.18$\pm$ 0.25 & 14.06$\pm$ 1.15 & 20.00 & 0.47$\pm$   0.32  \\
 & 54402 & 0.65$\pm$ 0.00 & 5.04$\pm$ 0.29 & 3.04$\pm$ 0.26 & 9.50$\pm$ 0.42 & 16.12$\pm$ 3.41 & 11.73$\pm$ 3.23 & 30.64$\pm$ 2.21 & 51.00 & 1.43$\pm$   0.99 \\
 & 54986 & 0.16$\pm$ 0.01 & 16.96$\pm$ 3.71 & 12.69$\pm$ 3.82 & 28.79$\pm$ 4.97 & 6.06$\pm$ 1.29 & 4.45$\pm$ 1.36 & 10.34$\pm$ 2.71 & 62.00 & 0.31$\pm$   0.22 \\
 & 55265 & 0.57$\pm$ 0.01 & 1.24$\pm$ 0.08 & 0.58$\pm$ 0.05 & 2.92$\pm$ 0.20 & 3.67$\pm$ 0.33 & 2.67$\pm$ 0.32 & 6.88$\pm$ 0.61 & 14.00 & 0.32$\pm$   0.22 \\
 & 55669 & 0.71$\pm$ 0.01 & 10.14$\pm$ 2.93 & 10.83$\pm$ 3.63 & 5.88$\pm$ 2.45 & 31.59$\pm$ 3.23 & 30.68$\pm$ 4.59 & 1.71$\pm$ 0.29 & 97.00 & 2.10$\pm$   1.45 \\
XTE J1701$-$462 & 53767 & 0.91$\pm$ 0.01 & 11.95$\pm$ 2.26 & 7.86$\pm$ 1.97 & 24.88$\pm$ 1.55 & 220.68$\pm$ 13.41 & 295.43$\pm$ 25.99 & 39.86$\pm$ 2.03 & 566.00 & 33.06$\pm$   9.77  \\
RX J1709.5$-$2639 & 50456 & 0.22$\pm$ 0.01 & 1.66$\pm$ 0.87 & \nodata & \nodata & 31.97$\pm$ 7.19 & 25.94$\pm$ 6.13 & 52.75$\pm$ 33.75 & 77.00 & 1.69$\pm$   0.68 \\
 & 52306 & 0.18$\pm$ 0.02 & 32.34$\pm$ 24.38 & 43.52$\pm$ 47.34 & 7.12$\pm$ 1.91 & 11.53$\pm$ 3.08 & 5.21$\pm$ 2.00 & 29.27$\pm$ 13.82 & 102.00 & 1.92$\pm$   0.77 \\
 & 54431 & 0.26$\pm$ 0.03 & 7.24$\pm$ 1.98 & 6.83$\pm$ 3.03 & 7.92$\pm$ 1.74 & 27.06$\pm$ 9.57 & 13.78$\pm$ 7.40 & 55.06$\pm$ 18.43 & 73.00 & 1.97$\pm$   0.79 \\
XTE J1723$-$376 & 51189 & 0.16$\pm$ 0.02 & 22.74$\pm$ 11.25 & 32.81$\pm$ 23.96 & 1.34$\pm$ 0.30 & 45.19$\pm$ 4.60 & 21.42$\pm$ 5.54 & 12.18$\pm$ 5.63 & 147.00 & 1.27$\pm$   0.76 \\
MXB 1730$-$33 & 50188 & 0.26$\pm$ 0.01 & 0.78$\pm$ 0.15 & 0.55$\pm$ 0.11 & 4.38$\pm$ 0.89 & 8.40$\pm$ 1.16 & 9.28$\pm$ 2.01 & 6.74$\pm$ 0.82 & 22.00 & 0.44$\pm$   0.30 \\
 & 50387 & 0.30$\pm$ 0.01 & \nodata & \nodata & 2.37$\pm$ 0.33 & 11.81$\pm$ 3.95 & 12.41$\pm$ 5.49 & 9.97$\pm$ 1.09 & \nodata & \nodata \\
 & 50625 & 0.38$\pm$ 0.01 & 2.33$\pm$ 0.42 & \nodata & \nodata & 7.65$\pm$ 0.79 & 9.46$\pm$ 1.49 & 4.18$\pm$ 0.28 & 24.00 & 0.47$\pm$   0.32 \\
 & 50843 & 0.27$\pm$ 0.01 & 5.71$\pm$ 1.23 & \nodata & \nodata & 9.20$\pm$ 0.75 & 10.75$\pm$ 1.27 & 5.19$\pm$ 0.70 & 32.00 & 0.50$\pm$   0.34 \\
 & 51047 & 0.16$\pm$ 0.02 & 1.49$\pm$ 1.58 & 1.64$\pm$ 2.30 & 1.02$\pm$ 0.22 & 7.39$\pm$ 1.13 & 10.32$\pm$ 2.20 & 1.68$\pm$ 0.84 & 20.00 & 0.35$\pm$   0.24 \\
 & 51250 & 0.28$\pm$ 0.01 & 3.42$\pm$ 0.69 & 3.90$\pm$ 1.00 & 1.37$\pm$ 0.28 & 6.96$\pm$ 0.55 & 8.89$\pm$ 1.06 & 2.88$\pm$ 0.63 & 30.00 & 0.35$\pm$   0.24 \\
 & 51453 & 0.24$\pm$ 0.02 & 1.99$\pm$ 0.26 & \nodata & \nodata & 9.72$\pm$ 3.02 & 9.18$\pm$ 3.51 & 12.19$\pm$ 3.83 & 30.00 & 0.42$\pm$   0.29 \\
 & 51739 & 0.15$\pm$ 0.01 & 3.03$\pm$ 1.02 & 1.04$\pm$ 0.74 & 5.80$\pm$ 3.35 & 6.03$\pm$ 1.16 & 5.58$\pm$ 1.56 & 7.08$\pm$ 1.11 & 19.00 & 0.21$\pm$   0.14 \\
 & 51953 & 0.17$\pm$ 0.01 & 0.85$\pm$ 0.49 & \nodata & \nodata & 7.66$\pm$ 1.09 & 8.94$\pm$ 1.69 & 3.44$\pm$ 0.44 & 22.00 & 0.31$\pm$   0.21 \\
 & 52051 & 0.16$\pm$ 0.01 & 1.32$\pm$ 0.36 & 0.82$\pm$ 0.28 & 3.51$\pm$ 0.81 & 9.04$\pm$ 1.66 & 8.94$\pm$ 2.45 & 9.27$\pm$ 1.70 & 24.00 & 0.29$\pm$   0.20 \\
 & 52387 & 0.14$\pm$ 0.01 & 2.93$\pm$ 0.48 & 0.53$\pm$ 0.11 & 11.97$\pm$ 1.65 & 4.49$\pm$ 3.21 & 4.36$\pm$ 4.17 & 4.88$\pm$ 0.79 & 21.00 & 0.26$\pm$   0.18 \\
 & 53103 & 0.17$\pm$ 0.01 & 2.35$\pm$ 0.53 & 2.45$\pm$ 0.66 & 1.89$\pm$ 0.64 & 7.80$\pm$ 3.46 & 7.73$\pm$ 4.55 & 8.02$\pm$ 1.46 & 26.00 & 0.33$\pm$   0.23 \\
 & 53602 & 0.13$\pm$ 0.02 & 4.84$\pm$ 0.88 & 0.74$\pm$ 0.19 & 2.76$\pm$ 0.52 & 6.52$\pm$ 0.89 & 6.15$\pm$ 1.35 & 7.20$\pm$ 1.19 & 22.00 & 0.29$\pm$   0.20 \\
 & 53749 & 0.22$\pm$ 0.02 & 10.70$\pm$ 4.51 & \nodata & \nodata & 3.54$\pm$ 0.77 & 3.80$\pm$ 1.01 & 2.39$\pm$ 0.90 & 31.00 & 0.46$\pm$   0.31 \\
 & 53907 & 0.21$\pm$ 0.02 & 0.80$\pm$ 0.20 & 0.48$\pm$ 0.14 & 2.46$\pm$ 0.70 & 9.66$\pm$ 2.35 & 11.94$\pm$ 3.96 & 3.43$\pm$ 0.72 & 23.00 & 0.31$\pm$   0.22  \\
 & 54265 & 0.16$\pm$ 0.01 & 1.97$\pm$ 0.69 & \nodata & \nodata & 9.56$\pm$ 3.06 & 7.96$\pm$ 4.58 & 10.45$\pm$ 2.76 & 28.00 & 0.35$\pm$   0.24 \\
 & 54430 & 0.21$\pm$ 0.01 & 1.35$\pm$ 0.39 & 1.25$\pm$ 0.45 & 1.78$\pm$ 0.25 & 7.21$\pm$ 4.29 & 9.74$\pm$ 8.38 & 1.47$\pm$ 0.29 & 20.00 & 0.24$\pm$   0.17  \\
 & 54734 & 0.13$\pm$ 0.02 & 4.39$\pm$ 0.96 & 3.41$\pm$ 1.06 & 6.82$\pm$ 1.29 & 4.36$\pm$ 1.71 & 3.88$\pm$ 2.25 & 5.58$\pm$ 2.76 & 21.00 & 0.22$\pm$   0.15 \\
 & 54842 & 0.27$\pm$ 0.01 & 1.60$\pm$ 1.73 & \nodata & \nodata & 4.04$\pm$ 1.58 & 4.14$\pm$ 2.09 & 3.68$\pm$ 0.95 & 14.00 & 0.30$\pm$   0.20 \\
 & 54976 & 0.14$\pm$ 0.01 & 0.89$\pm$ 0.34 & 0.40$\pm$ 0.20 & 2.30$\pm$ 0.41 & 5.85$\pm$ 4.09 & 7.73$\pm$ 7.70 & 1.40$\pm$ 0.27 & 17.00 & 0.16$\pm$   0.11 \\
 & 55074 & 0.18$\pm$ 0.02 & 1.15$\pm$ 0.20 & 0.80$\pm$ 0.20 & 2.06$\pm$ 0.49 & 6.63$\pm$ 1.70 & 4.11$\pm$ 1.53 & 12.66$\pm$ 2.88 & 18.00 & 0.23$\pm$   0.16 \\
 & 55220 & 0.22$\pm$ 0.02 & 18.67$\pm$ 9.28 & 12.98$\pm$ 9.61 & 33.29$\pm$ 23.84 & 13.01$\pm$ 3.89 & 6.77$\pm$ 2.94 & 4.16$\pm$ 1.25 & 76.00 & 0.77$\pm$   0.53 \\
 & 55395 & 0.19$\pm$ 0.01 & 0.89$\pm$ 0.35 & 0.79$\pm$ 0.58 & 1.02$\pm$ 0.30 & 8.28$\pm$ 4.43 & 3.69$\pm$ 2.56 & 25.09$\pm$ 13.76 & 22.00 & 0.26$\pm$   0.18 \\
 XTE J1739$-$285 & 51475 & 0.16$\pm$ 0.01 & 0.98$\pm$ 0.33 & \nodata & \nodata & 40.22$\pm$ 10.22 & 42.80$\pm$ 15.67 & 15.34$\pm$ 6.12 & 86.00 & 0.55$\pm$   0.28  \\
GRO J1744$-$28 & 50094 & 1.29$\pm$ 0.02 & \nodata & \nodata & \nodata & 42.18$\pm$ 1.57 & 28.11$\pm$ 1.52 & 72.99$\pm$ 2.57 & NaN & \nodata \\
 & 50474 & 0.81$\pm$ 0.02 & 19.78$\pm$ 1.69 & 18.09$\pm$ 2.20 & 24.50$\pm$ 2.84 & 28.19$\pm$ 1.42 & 20.40$\pm$ 1.39 & 50.66$\pm$ 4.07 & 102.00 & 6.50$\pm$   0.77 \\
IGR J17473$-$2721 & 53347 & 0.15$\pm$ 0.02 & 4.65$\pm$ 2.97 & 5.32$\pm$ 4.49 & 2.30$\pm$ 1.59 & 16.70$\pm$ 13.06 & \nodata & \nodata & 56.00 & 0.13$\pm$   0.05 \\
 & 54642 & 0.42$\pm$ 0.02 & 29.42$\pm$ 3.41 & 41.67$\pm$ 7.17 & 3.09$\pm$ 0.47 & 14.19$\pm$ 3.33 & 8.78$\pm$ 2.61 & 33.40$\pm$ 8.01 & 95.00 & 1.38$\pm$   0.41 \\ 
EXO 1745$-$248 & 51776 & 0.53$\pm$ 0.02 & 3.89$\pm$ 0.28 & 3.85$\pm$ 0.31 & 4.31$\pm$ 1.00 & 18.13$\pm$ 1.84 & 19.98$\pm$ 3.14 & 13.99$\pm$ 2.05 & 53.00 & 1.91$\pm$   1.32  \\
 & 52459 & 0.22$\pm$ 0.02 & 4.97$\pm$ 1.02 & 4.77$\pm$ 1.30 & 5.58$\pm$ 1.50 & 5.74$\pm$ 1.44 & 2.22$\pm$ 0.76 & 15.78$\pm$ 3.51 & 29.00 & 0.55$\pm$   0.38 \\
 & 55486 & 0.61$\pm$ 0.03 & 33.24$\pm$ 4.98 & 12.94$\pm$ 4.77 & 68.23$\pm$ 33.77 & 40.55$\pm$ 11.71 & 47.50$\pm$ 21.89 & 28.54$\pm$ 4.18 & 166.00 & 7.34$\pm$   5.06\\ 
1A 1744$-$361 & 52955 & 0.15$\pm$ 0.01 & 2.53$\pm$ 2.45 & 0.63$\pm$ 0.68 & 19.30$\pm$ 4.95 & 29.58$\pm$ 26.49 & 9.12$\pm$ 10.94 & 91.63$\pm$ 45.22 & 82.00 & 0.41$\pm$   0.41 \\
4U 1745$-$203 & 52179 & 0.18$\pm$ 0.01 & 6.67$\pm$ 0.82 & 5.12$\pm$ 1.08 & 9.58$\pm$ 1.79 & 8.84$\pm$ 2.19 & 3.11$\pm$ 1.00 & 28.54$\pm$ 9.56 & 46.00 & 0.72$\pm$   0.07 \\
 & 53526 & 0.32$\pm$ 0.02 & 8.41$\pm$ 1.94 & 5.31$\pm$ 1.71 & 16.11$\pm$ 2.44 & 17.29$\pm$ 3.24 & 20.28$\pm$ 4.49 & 8.36$\pm$ 3.55 & 54.00 & 1.25$\pm$   0.12  \\
SAX J1750.8$-$2900 & 51971 & 0.14$\pm$ 0.01 & 3.56$\pm$ 1.16 & 3.98$\pm$ 2.12 & 2.81$\pm$ 0.86 & 14.79$\pm$ 4.94 & 19.17$\pm$ 9.37 & 4.61$\pm$ 1.36 & 39.00 & 0.16$\pm$   0.02 \\
 & 54588 & 0.17$\pm$ 0.02 & 7.63$\pm$ 1.70 & 9.90$\pm$ 3.30 & 2.80$\pm$ 0.68 & 20.80$\pm$ 8.50 & 18.89$\pm$ 10.93 & 25.38$\pm$ 5.83 & 69.00 & 0.51$\pm$   0.03 \\
2S 1803$-$245 & 50933 & 0.75$\pm$ 0.02 & 4.87$\pm$ 0.14 & 3.21$\pm$ 0.16 & 8.58$\pm$ 0.74 & 16.24$\pm$ 0.78 & 12.43$\pm$ 0.82 & 26.00$\pm$ 1.61 & 51.00 & 0.98$\pm$   0.79 \\
Aql X-1 & 50481 & 0.38$\pm$ 0.01 & 3.56$\pm$ 0.44 & 2.39$\pm$ 0.46 & 5.60$\pm$ 0.33 & 11.97$\pm$ 0.68 & 6.24$\pm$ 0.50 & 26.67$\pm$ 1.68 & 35.00 & 0.45$\pm$   0.18 \\
 & 50683 & 0.24$\pm$ 0.00 & 7.93$\pm$ 2.15 & 5.34$\pm$ 1.95 & 15.41$\pm$ 1.57 & 13.12$\pm$ 1.68 & 5.73$\pm$ 1.05 & 14.04$\pm$ 1.69 & 51.00 & 0.37$\pm$   0.15 \\
 & 50901 & 0.47$\pm$ 0.00 & 6.95$\pm$ 0.34 & 2.90$\pm$ 0.20 & 19.13$\pm$ 1.75 & 11.27$\pm$ 0.67 & 5.19$\pm$ 0.40 & 32.98$\pm$ 1.30 & 57.00 & 1.05$\pm$   0.42 \\
 & 51319 & 0.39$\pm$ 0.01 & 2.50$\pm$ 0.17 & \nodata & \nodata & 7.67$\pm$ 0.39 & 3.75$\pm$ 0.30 & 16.10$\pm$ 1.21 & 22.00 & 0.25$\pm$   0.10 \\
 & 51833 & 0.68$\pm$ 0.02 & 2.92$\pm$ 0.05 & 3.10$\pm$ 0.07 & 2.27$\pm$ 0.09 & 11.97$\pm$ 0.32 & 7.17$\pm$ 0.26 & 25.24$\pm$ 1.22 & 47.00 & 1.13$\pm$   0.45 \\
 & 52335 & 0.21$\pm$ 0.00 & 6.82$\pm$ 1.16 & 3.37$\pm$ 0.73 & 21.48$\pm$ 6.31 & 9.04$\pm$ 1.40 & 3.73$\pm$ 0.81 & 23.22$\pm$ 2.70 & 38.00 & 0.26$\pm$   0.10 \\
 & 52707 & 0.67$\pm$ 0.01 & 2.84$\pm$ 0.14 & 2.79$\pm$ 0.31 & 2.97$\pm$ 0.52 & 9.90$\pm$ 0.32 & 5.04$\pm$ 0.21 & 26.07$\pm$ 1.53 & 40.00 & 1.00$\pm$   0.40 \\
 & 53167 & 0.28$\pm$ 0.02 & 14.47$\pm$ 1.28 & 22.20$\pm$ 3.24 & 3.23$\pm$ 0.27 & 2.92$\pm$ 0.84 & 2.84$\pm$ 1.02 & 3.27$\pm$ 0.60 & 42.00 & 0.18$\pm$   0.07 \\
 & 54365 & 0.26$\pm$ 0.00 & 4.83$\pm$ 0.41 & 3.53$\pm$ 0.46 & 7.64$\pm$ 0.75 & 3.94$\pm$ 0.25 & 2.79$\pm$ 0.26 & 6.66$\pm$ 0.34 & 26.00 & 0.22$\pm$   0.09 \\
 & 55160 & 0.30$\pm$ 0.01 & 6.50$\pm$ 1.26 & 5.18$\pm$ 1.40 & 9.85$\pm$ 0.98 & 2.92$\pm$ 1.13 & 2.71$\pm$ 1.36 & 3.63$\pm$ 0.61 & 23.00 & 0.19$\pm$   0.08 \\
 & 55454 & 0.40$\pm$ 0.02 & 3.64$\pm$ 0.52 & \nodata & \nodata & 11.90$\pm$ 1.97 & 6.49$\pm$ 1.36 & 32.86$\pm$ 3.69 & 42.00 & 0.35$\pm$   0.14 \\

\enddata

 \tablenotetext{a}{The date when the outburst reaches its peak flux in the ASM light curve.}
\end{deluxetable}
\clearpage
\end{landscape}

%% file: ltab1.tex
\clearpage
\LongTables
\begin{landscape}

\begin{deluxetable}{lcccccccccc}
\tabletypesize{\tiny}
\centering
\tablecaption{The Parameters of Outbursts of Transient ULXs in M31 \label{m31}}
\tablehead{
\colhead{Source} 
&\colhead{$t_\mathrm{peak}$}
&\colhead{$L_\mathrm{peak}$ }
&\colhead{$\tau_\mathrm{rise,10\%-90\%}$} 
&\colhead{$\tau_\mathrm{rise,10\%-50\%}$} 
&\colhead{$\tau_\mathrm{rise,50\%-90\%}$} 
&\colhead{$\tau_\mathrm{decay,10\%-90\%}$}
&\colhead{$\tau_\mathrm{decay,10\%-50\%}$} 
&\colhead{$\tau_\mathrm{decay,50\%-90\%}$}
&\colhead{Outburst Duration} 
&\colhead{$E$} 
\\
\colhead{} 
&\colhead{MJD}
&\colhead{($10^{39}$ ergs s$^{-1}$)}
&\colhead{(days)} 
&\colhead{(days)} 
&\colhead{(days)} 
&\colhead{(days)}
&\colhead{(days)} 
&\colhead{(days)}
&\colhead{(days)} 
&\colhead{($10^{44}$ ergs)}
}

\startdata
XMMU J004243.61+412519 & 55957 & 1.26$\pm$ 0.02 & 8.69$\pm$ 0.48 & \nodata & \nodata & 75.57$\pm$ 3.05 & 53.31$\pm$ 3.91 & 105.80$\pm$ 6.91 & 195.00 & 112.50$\pm$ 0.43 \\
 
CXOM31 J004253.1+411422 & 55182 & $>$ 3.77$\pm$ 0.04 & \nodata & \nodata & \nodata & 33.05$\pm$ 6.57 & 38.63$\pm$ 9.81 & 12.84$\pm$ 1.33 & \nodata & \nodata \\
\enddata
\end{deluxetable}
\clearpage
 \end{landscape}